\documentclass[12pt]{article}
\usepackage{axodraw}
\newcommand{\be}{\begin{equation}}
\newcommand{\ee}{\end{equation}}
\newcommand{\sss}[1]{\mbox{\scriptsize #1}}

\newcommand{\LL}{{\cal L}}

\newcommand{\SL}{{\cal S}_{\sss{L}}}
\newcommand{\SNL}{{\cal S}_{\sss{NL}}}

\newcommand{\OO}{{\cal O}}

\newcommand{\A}{{\cal A}}

\newcommand{\X}{{\cal X}}

\newcommand{\T}{\mbox{T}}
\newcommand{\F}{\mbox{F}}
\newcommand{\Tr}{\mbox{Tr}\,}
\newcommand{\G}{\mbox{G}}
\newcommand{\TF}{\mbox{\boldmath $F$}}
\newcommand{\TA}{\mbox{\boldmath $A$}}
\newcommand{\TW}{\mbox{\boldmath $W$}}

\newcommand{\ps}{p\hspace{-0.42em}/}

\newcommand{\partials}{\partial\hspace{-0.50em}/\hspace{0.1em}}
\newcommand{\Ds}{D\hspace{-0.63em}/\hspace{0.1em}}
\newcommand{\TAs}{\TA\hspace{-0.65em}/\hspace{0.1em}}

\begin{document}
\pagestyle{empty}
\begin{flushright}
  DTP/99/96 \\
  INLO-PUB 18/99
\end{flushright}
\vspace*{5mm}
\begin{center}
  {\bf AN EFFECTIVE LAGRANGIAN APPROACH FOR UNSTABLE PARTICLES} \\
  \vspace*{1cm} 
  {\bf W.~Beenakker}$^{*)}$\\ 
  \vspace{0.3cm}
  Physics Department, University of Durham, Durham DH1 3LE, England\\
  \vspace{0.5cm}
  {\bf F.A.~Berends} \ \  
  {\bf and}  \ \ 
  {\bf A.P.~Chapovsky}$^{\dagger)}$\\
  \vspace{0.3cm}
  Instituut--Lorentz, University of Leiden, The Netherlands\\
  \vspace*{2cm}  
  {\bf ABSTRACT} \\ 
\end{center}
\vspace*{5mm}
\noindent
We propose a novel procedure for handling processes that involve unstable 
intermediate particles. By using gauge-invariant effective Lagrangians it is 
possible to perform a gauge-invariant resummation of (arbitrary) self-energy 
effects. For instance, gauge-invariant tree-level amplitudes can be constructed
with the decay widths of the unstable particles properly included in the 
propagators. In these tree-level amplitudes modified vertices are used, which 
contain extra gauge-restoring terms prescribed by the effective Lagrangians.
We discuss the treatment of the phenomenologically important 
unstable particles, like the top-quark, the $W$- and $Z$-bosons, and the 
Higgs-boson, and derive the relevant modified Feynman rules explicitly.
\vspace*{3cm}\\ 
\begin{flushleft}
  September 1999
\end{flushleft}
\noindent 
\rule[.1in]{16.5cm}{.002in}

\noindent
$^{*)}$Research supported by a PPARC Research Fellowship.\\
$^{\dagger)}$Research supported by the Stichting FOM.
\vspace*{0.3cm}

\vfill\eject

\setcounter{page}{1}
\pagestyle{plain}


\section{Introduction}
\label{sec:intro}

Many of the interesting reactions at present-day and future collider 
experiments involve a multitude of unstable particles during the intermediate 
stages. In view of the high precision of the experiments, the proper treatment 
of these unstable particles has become a demanding exercise, since on-shell 
approximations are simply not good enough anymore. A proper treatment of 
unstable particles requires the resummation of the corresponding self-energies 
to all orders. In this way the singularities originating from the poles in the 
on-shell propagators are regularized by the imaginary parts contained in the
self-energies, which are closely related to the decay widths of the unstable 
particles. The perturbative resummation itself involves a simple geometric 
series and is therefore easy to perform. However, this simple procedure 
harbours the serious risk of breaking gauge invariance. Gauge invariance is 
guaranteed order by order in perturbation theory. Unfortunately one takes into 
account only part of the higher-order terms by resumming the self-energies. 
This results in a mixing of different orders of perturbation theory and 
thereby jeopardizes gauge invariance, even if the self-energies themselves are 
extracted in a gauge-invarant way.

During recent years awareness has been raised regarding the seriousness of the
problem. It was shown explicitly how these gauge-breaking effects, which are 
formally of higher order in the expansion parameter, can nevertheless have 
profound repercussions on physical observables~\cite{west,lapp-kek,fls1,fls2}. 
This applies in particular to kinematical situations that 
approach asymptotic limits, like space-like virtual photons close to the 
on-shell limit~\cite{west,lapp-kek,fls1} or longitudinal gauge bosons at high 
energies~\cite{fls2}. These asymptotic regimes are characterized by strong 
gauge cancellations, which are governed by the Ward identities of the theory. 
Any small gauge-breaking effect can upset these intricate gauge cancellations 
and can therefore be amplified significantly.

A solution to the problem is provided by the so-called 
pole-scheme~\cite{pole-scheme}, which allows the gauge-invariant calculation 
of matrix elements in the presence of unstable particles. The pole-scheme 
amounts to a systematic expansion of the matrix elements around the complex 
poles in the unstable-particle propagators. This can be viewed as a 
prescription for performing an effective expansion in powers of $\Gamma_i/M_i$,
where $M_i$ and $\Gamma_i$ stand for the masses and widths of the unstable 
particles. The residues in the pole expansion are physically observable and 
therefore gauge-invariant. In reactions with multiple unstable-particle 
resonances it is rather awkward to perform the complete pole-scheme expansion
with all its subtleties in the treatment of the off-shell phase 
space~\cite{dpa-ww}. Therefore one usually approximates the expansion by 
retaining only the terms with the highest degree of resonance. 
This approximation is called the leading-pole approximation. The accuracy of 
the approximation is typically $\OO(\Gamma_i/M_i)$, making it a suitable tool 
for calculating radiative corrections, since in that case the errors are 
further suppressed by powers of the coupling constant~\cite{dpa-ww}. 
The errors induced at the lowest-order level, however, are as large as the 
radiative corrections themselves. In view of the high precision of present-day 
collider experiments, this is not acceptable. Therefore either the lowest-order
expansion has to be performed explicitly or an alternative gauge-invariant 
resummation method should be used. 

A few years ago a dedicated method was developed for the gauge-invariant 
treatment of unstable gauge bosons~\cite{fls1,fls2,fls-baur,fls-others}. 
This so-called fermion-loop scheme exploits the fact that the unstable gauge 
bosons decay exclusively into fermions (at lowest order). Based on this 
observation, it proved natural to resum the fermionic one-loop self-energies 
and include all other possible one-particle-irreducible fermionic one-loop 
corrections. This resummation of one-particle-irreducible fermionic one-loop 
corrections involves the closed subset of all $\OO([N_c^f\alpha/\pi]^n)$ 
contributions (with $N_c^f$ denoting the colour degeneracy of fermion $f$), 
which makes it manifestly gauge-invariant. Unfortunately this method does not 
work for particles that also have bosonic decay modes. Moreover, the inclusion 
of a full-fledged set of one-loop corrections in a lowest-order amplitude 
tends to overcomplicate things.

A more general and rapidly developing method is the so-called pinch technique
(PT)~\cite{pinch}. This method can be viewed as an extension of the 
fermion-loop scheme to the bosonic sector. It amounts to a re-organization of 
the various one-loop Green's functions in terms of gauge-invariant off-shell 
building blocks, labelled by the kinematical characteristics of the terms that 
are included (e.g~self-energy-like terms, vertex-like terms, etc.). These 
building blocks satisfy ghost-free tree-level Ward identities (like in the 
fermion-loop scheme) and can be combined into gauge-invariant amplitudes. 
All this is achieved by making explicit use of the full Ward identities of the
theory. After having applied the pinching procedure, the one-loop PT 
self-energies can be resummed in the resonant amplitudes~\cite{pinch-resum}. 
The gauge invariance of this resummation then follows from the tree-level-like 
Ward identities of the non-resummed (vertex-like, box-like, etc.) one-loop 
corrections (see Ref.~\cite{bkg} for a formal proof based on the 
background-field-method). The PT is therefore a suitable candidate for treating
lowest-order reactions involving unstable particles, although the lowest-order
amplitudes will be quite complicated in view of the full set of non-resummed 
one-loop corrections. The complexity of these mandatory non-resummed one-loop
corrections grows strongly with the amount of final-state particles in the 
lowest-order reaction, just like in the fermion-loop scheme. The terminology 
`lowest-order' refers to the fact that resonant amplitudes are dominated by 
the decay widths in the propagators, which are calculated in lowest order in 
the PT. In order to go beyond the lowest order, the imaginary parts of the 
two-loop self-energies are needed. Gauge invariance of the resummation 
procedure in turn requires the inclusion of the relevant imaginary parts of 
the other two-loop corrections. At the moment some first attempts are under 
way to extend the PT beyond the one-loop order~\cite{pinch-two-loop}. However, 
there is still a long way to go. Developing a fundamental (non-diagrammatic) 
understanding of the PT might be the most important next step in this context.

So, the need remains for a novel, preferably non-diagrammatic method to solve 
the full set of Ward identities. Ideally 
speaking, such an alternative method should be applicable to arbitrary 
reactions, involving all possible unstable particles and an unspecified 
amount of stable external particles. At the same time the 
gauge-restoring terms should be kept to a minimum. In this paper we propose 
such a novel technique for tree-level processes. By using gauge-invariant 
non-local effective Lagrangians, it is possible to generate the self-energy
effects in the propagators as well as the required gauge-restoring terms in 
the multi-particle interactions. These multi-particle interactions can be 
derived explicitly in a relatively concise form.
Of course one should generate physically sensible self-energies, so that 
the tree-level calculations are phenomenologically meaningful.

The paper is organized as follows. In Sect.~\ref{sec:fls} we briefly discuss
the essence of the gauge-invariance problem. The non-local effective-Lagrangian
method for the resummation of self-energies is introduced in 
Sects.~\ref{sec:top} and \ref{sec:ww} within the framework of an unbroken 
non-abelian $SU(N)$ gauge theory with fermions. In Sect.~\ref{sec:sm} we apply 
the method to the treatment of unstable particles in the Standard Model and 
give some simple examples. In the appendices we list some useful non-local 
Feynman rules, thus demonstrating the application of the method.


\section{The gauge-invariance problem: a simple example}
\label{sec:fls}

In this section we show the origin of the gauge-invariance problem associated
with the resummation of self-energies, which is a minimal requirement for 
treating unstable particles. To this end we consider the simple example of an 
unbroken non-abelian $SU(N)$ gauge theory with fermions and subsequently 
integrate out these fermions. This example also allows to make a direct link 
to the philosophy behind the fermion-loop scheme.

First we fix our notation and introduce some conventions, which will be used 
throughout this paper. The $SU(N)$ generators in the fundamental representation
are denoted by $\T^a$ with $\,a=1,\cdots,N^2\!-\!1$. They are normalized 
according to $\Tr(\T^a\T^b)=\delta^{ab}/2$ and obey the commutation relation
$\Bigl[ \T^a,\T^b \Bigr] = if^{abc}\,\T^c$. In the adjoint representation the 
generators $\F^a$ are given by $(\F^a)^{bc} = -if^{abc}$. The Lagrangian of 
the unbroken $SU(N)$ gauge theory with fermions can be written as 
\be
\label{lagr}
  \LL(x) 
  = 
  -\,\frac{1}{2}\,\Tr\Bigl[ \TF_{\mu\nu}(x)\,\TF^{\mu\nu}(x) \Bigr] 
  + \bar{\psi}(x)\, (i\,\Ds - m)\, \psi(x),
\ee
with
\be
  \TF_{\mu\nu} \equiv \T^a F_{\mu\nu}^a = \frac{i}{g}\,[D_{\mu},D_{\nu}], 
  \qquad 
  D_{\mu} = \partial_{\mu} - ig\,\T^a A_{\mu}^a 
            \equiv \partial_{\mu} - ig\TA_{\mu}.
\ee
Here $\psi$ is a fermionic $N$-plet in the fundamental representation of
$SU(N)$ and $A_{\mu}^a$ are the ($N^2\!-\!1$) non-abelian $SU(N)$ gauge fields,
which form a multiplet in the adjoint representation. 
The Lagrangian (\ref{lagr}) is invariant under the 
$SU(N)$ gauge transformations
$$
  \psi(x) \to \psi'(x) = \G(x)\,\psi(x), 
$$
\be
\label{gauge_trans}
  \TA_{\mu}(x) \to \TA'_{\mu}(x) = \G(x)\,\TA_{\mu}(x)\,\G^{-1}(x)
           + \frac{i}{g}\,\G(x)\Bigl[ \partial_{\mu}\G^{-1}(x) \Bigr],
\ee
with the $SU(N)$ group element defined as $\G(x)=\exp[ig\,\T^a\theta^a(x)]$. 
The covariant derivative $D_{\mu}$ and field strength $\TF_{\mu\nu}$ both 
transform in the adjoint representation
\be
  D_{\mu} \to \G(x)\,D_{\mu}\,\G^{-1}(x), 
  \qquad 
  \TF_{\mu\nu}(x) \to \G(x)\,\TF_{\mu\nu}(x)\,\G^{-1}(x).
\ee
Since the Lagrangian is quadratic in the fermion fields, one can integrate 
them out exactly in the functional integral. The resulting effective action is 
then given by
\be
  i\,S_{\sss{eff}}[J] 
  = 
  i\int d^4x\,\biggl\{ 
            -\,\frac{1}{2}\,\Tr\Bigl[ \TF_{\mu\nu}(x)\,\TF^{\mu\nu}(x) \Bigr]
            + J_{\mu}^a(x)\,A^{a,\,\mu}(x) \biggr\}
  + \Tr\Bigl[ \ln(-\,\Ds - i\,m) \Bigr],            
\ee
with $J_{\mu}^a(x)$ denoting the gauge-field sources. The trace on the 
right-hand side has to be taken in group, spinor, and coordinate space.
As a next step one can expand the effective action in terms of the coupling 
constant
\begin{eqnarray}
\label{fls/expansion}
  \Tr\Bigl[ \ln(-\,\Ds - i\,m) \Bigr] &=& 
            \Tr\Bigl[ \ln(-\,\partials - i\,m) \Bigr] 
          + \Tr\biggl[ \ln\Bigl( 1+\frac{g}{i\,\partials-m}\,\TAs 
                          \Bigr) \biggr]
  \nonumber \\[1mm]
                                             &=& 
            \Tr\Bigl[ \ln(-\,\partials - i\,m) \Bigr]
          + \sum\limits_{n=1}^{\infty} \frac{(-1)^{n-1}}{n}\,
            \Tr\Biggl[ \biggl( \frac{g}{i\,\partials-m}\,\TAs
                       \biggr)^n \Biggr].
\end{eqnarray}
Note that the left-hand side of Eq.~(\ref{fls/expansion}) is gauge-invariant as
a result of the trace-log operation. In contrast, the separate terms of the 
expansion on the right-hand side are not gauge-invariant. This is due to the 
fact that, unlike in the abelian case, the non-abelian gauge transformation 
(\ref{gauge_trans}) mixes different powers of the gauge field $A_{\mu}$ in 
Eq.~(\ref{fls/expansion}). 
Thus, if one truncates the series on the 
right-hand side of Eq.~(\ref{fls/expansion}) one will in general break gauge 
invariance. From Eq.~(\ref{fls/expansion}) it is also clear that the fermionic 
part of the effective action induces higher-order interactions between the 
gauge bosons. 

What are these higher-order interactions? Let us consider the quadratic
gauge-field contribution 
\be
\label{fls/AA} 
  -\,\frac{1}{2}\,\Tr\Biggl[ \biggl( \frac{g}{i\,\partials-m}\,\TAs
                             \biggr)^2 \Biggr]
  =
  -\,\frac{1}{2}\,\int d^4x\,d^4y\,\Tr\Bigl[ O(x,y)\,O(y,x) \Bigr],
\ee
where
\be 
  O(x,y) = g\,S^{(0)}_{\sss{F}}(x-y)\,\TAs(y)
\ee
and $i\,S^{(0)}_{\sss{F}}(x-y)=\,<\!0\,|\,T(\psi(x)\,\bar{\psi}(y))\,
|\,0\!>_{\sss{free}}\,$ is the free fermion propagator. The trace on the 
right-hand side of Eq.~(\ref{fls/AA})
has to be taken in group and spinor space. A quick glance at this quadratic 
gauge-field contribution reveals that it is just the one-loop self-energy of 
the gauge boson induced by a fermion loop. In the same way, the higher-order 
terms $\sim g^n A^n$ in Eq.~(\ref{fls/expansion}) are just the fermion-loop 
contributions to the $n$-point gauge-boson vertices.

One can truncate the expansion in Eq.~(\ref{fls/expansion}) at $n=2$, thus 
taking into account only the gauge-boson self-energy term and neglecting the 
fermion-loop contributions to the higher-point gauge-boson vertices. This is
evidently the simplest procedure for performing the Dyson resummation of the 
fermion-loop self-energies. However, as was pointed out above, truncation of
Eq.~(\ref{fls/expansion}) at any finite order in $g$ in general breaks 
gauge invariance. This leads to the important observation that, {\it although 
the resummed fermion-loop self-energies are gauge-independent by themselves, 
the resummation is nevertheless responsible for gauge-breaking effects in the
higher-point gauge-boson interactions through its inherent mixed-order nature}.
Another way of understanding this is provided by the gauge-boson Ward
identities. Since the once-contracted $n$-point gauge-boson vertex can be 
expressed in terms of $(n\!-\!1)$-point vertices (see Sect.~\ref{sec:ww}), it 
is clear that gauge invariance is violated if the self-energies are resummed 
without adding the necessary compensating terms to the higher-point vertices.

An alternative is to keep all the terms in Eq.~(\ref{fls/expansion}). Then the 
matrix elements derived from the effective action will be gauge-invariant. 
Keeping all the terms means that we will have to take into account not only 
the fermion-loop self-energy in the propagator, but also all the possible 
fermion-loop contributions to the higher-point gauge-boson vertices. This is 
exactly the prescription of the fermion-loop scheme 
(FLS)~\cite{fls1,fls2,fls-baur}. 
Although the FLS guarantees gauge invariance of the matrix elements, it has 
disadvantages as well. Its general applicability is limited to those situations
where non-fermionic particles can effectively be discarded in the 
self-energies, as is for instance the case for $\Gamma_W$ and $\Gamma_Z$ at 
lowest order. Another disadvantage is that in the FLS one is forced to do the 
loop calculations, even when calculating lowest-order quantities. For example, 
the calculation of the tree-level matrix element for the process 
$e^+e^- \to 4f\gamma\,$ involves a four-point gauge-boson interaction, which 
has to be corrected by fermion loops in the FLS. This overcomplicates an 
otherwise lowest-order calculation.

It is clear that the FLS provides more than we actually need. It does not only 
provide gauge invariance for the Dyson resummed matrix elements at a given 
order in the coupling constant, but it also takes into account all
fermion-loop corrections at that given order. In the vicinity of unstable
particle resonances the imaginary parts of the fermion-loop self-energies are 
effectively enhanced by ${\cal O}(1/g^2)$ with respect to the other 
fermion-loop corrections. Therefore, what is really needed is only a minimal 
subset of the non-enhanced contributions such that gauge invariance
is restored. In a sense one is looking for a minimal solution of a system of 
Ward identities. The FLS provides a solution, but this solution is far from 
minimal and is only practical for particles that decay exclusively into 
fermions. Since the decay of unstable particles is a physical phenomenon,
it seems likely that there exists a simpler and more natural method for 
constructing a solution to a system of Ward identities, without an explicit
reference to fermions. In the following sections we will try to indicate how 
such a solution can be constructed.
Although the solution will be valid for arbitrary self-energies, in practical
calculations one will take physically sensible self-energies just like the FLS 
does.


\section{An effective-Lagrangian approach for fermions}
\label{sec:top}

In this section we propose a scheme that allows a gauge-invariant resummation
of fermion self-energies in tree-level processes, without having to resort
to a complete set of loop corrections. This scheme will form the basis for
the treatment of unstable fermions in the SM, like the top-quark. The crucial 
ingredient in the scheme is the use of non-local effective Lagrangians.
We start off by briefly repeating the non-local Lagrangian formalism of 
Ref.~\cite{nl-terning}, where the concept of non-local Lagrangians was used 
for a completely different purpose. 

The usual local $SU(N)$ gauge theory with fermions has a Lagrangian given by 
Eq.~(\ref{lagr}), which is invariant under the gauge transformations 
(\ref{gauge_trans}). For the gauging procedure of the non-local Lagrangians we 
will need one more ingredient, the {\it path-ordered exponential}, which is 
defined as 
\begin{eqnarray}
\label{Pexp}
  U(x,y) &=& U^{\dagger}(y,x) \ =\ \mbox{P}\exp\Biggl[ 
             -\,ig\int\limits_x^y\TA_{\mu}(\omega)\,d\omega^{\mu} \Biggr]
             \\
         &=& \lim_{d\omega_i \to 0}
             \biggl( 1 - ig\TA_{\mu}(x)\,d\omega_1^{\mu} \biggr)
             \biggl( 1 - ig\TA_{\mu}(x+d\omega_1)\,d\omega_2^{\mu} \biggr)
             \dots
             \biggl( 1 - ig\TA_{\mu}(y-d\omega_n)\,d\omega_n^{\mu} \biggr).
             \nonumber
\end{eqnarray}
Here $d\omega^{\mu}$ is the element of integration along some path 
$\Omega(x,y)$ that connects the points $x$ and $y$. In principle we are free
to choose this particular path. For reasons that will become clear from the 
discussion presented below, we will make our choice in such a way that
\be
\label{nullpath}
  \delta^{(4)}(x-y) \int \limits_x^y \TA_{\mu}(\omega)\,d\omega^{\mu} = 0
\ee
and
\be
\label{U/der}
  \partial^{\mu}_y \int\limits_x^y f(\omega)\,d\omega^{\nu}
  = 
  f(y)\,g^{\mu\nu}.
\ee
The first condition implies that the null path $\Omega(x,x)$ always has zero 
length, i.e.~it does not involve a closed loop. The second condition fixes the
properties of the path-ordered exponentials under differentiation.
The so-defined path-ordered exponential transforms as 
\be
  U(x,y) \to \G(x)\,U(x,y)\,\G^{-1}(y)
\ee
under the $SU(N)$ gauge transformations. It hence carries the gauge 
transformation from one space-time point to the other. With the help of the 
path-ordered exponential one can rewrite the local action corresponding to the 
fermionic part of the Lagrangian (\ref{lagr}) according to
\be
\label{lagr/2}
  \SL 
  = 
  \int d^4x\,d^4y\,\bar{\psi}(x)\,\delta^{(4)}(x-y)\,
  (i\,\partials_y-m)\,U(x,y)\,\psi(y).
\ee
Note that the action (\ref{lagr/2}) and the one obtained from Eq.~(\ref{lagr})
are only equivalent because of the condition (\ref{nullpath}).

Using this gauging procedure we can now add a gauge-invariant non-local term to
the Lagrangian:
\be
\label{lagr/nl}
  \LL_{\sss{NL}}(x,y) 
  = 
  \bar{\psi}(x)\,\Sigma_{\sss{NL}}(x-y)\,U(x,y)\,\psi(y).
\ee
It contains a non-local coefficient $\Sigma_{\sss{NL}}(x-y)$, which will play 
the role of a self-energy correction to the free fermion propagator. The 
argument of this coefficient is $x-y$ as a result of translational invariance. 
For calculational simplicity we will assume that $\Sigma_{\sss{NL}}(x-y)$ is 
mass-like, i.e.~it is diagonal in spinor space. Consequently it has to be a 
function of the scalar invariant $(x-y)^2$. This condition is not essential for
the method, but it will suit our purposes later on. With the new non-local 
term added to the Lagrangian, the total gauge-invariant fermionic action 
becomes
\begin{eqnarray}
  {\cal S}  &=& \SL + \SNL, \nonumber \\[1mm]
  \SL       &=& \int d^4x\,\bar{\psi}(x)\,\Bigl[ i\,\partials-m+g\,\TAs(x) 
                \Bigr]\,\psi(x), \nonumber \\
  \SNL      &=& \int d^4x\,d^4y\,\bar{\psi}(x)\,\Sigma_{\sss{NL}}(x-y)
                \,U(x,y)\,\psi(y).
\end{eqnarray}

As a next step we derive the relevant Feynman rules from the action ${\cal S}$.
First we verify that the non-local term acts as a self-energy correction to 
the free fermion propagator:
\be 
  \begin{picture}(60,15)(0,-2)
    \ArrowLine(0,0)(30,0)
    \ArrowLine(30,0)(60,0)
    \Text(12,7)[lb]{$p$}
    \Text(42,7)[lb]{$p'$}
    \GCirc(30,0){3}{0}
  \end{picture} 
  \ :\ i\,\Sigma(y,z) 
  = 
  \frac{i\,\delta^2 {\cal S}}{\delta\psi(y)\,\delta\bar{\psi}(z)}
  \Biggl.\Biggr|_{A=\psi=\bar{\psi}=0} 
  =
  -\,(\partials_z + i\,m)\,\delta^{(4)}(z-y) + i\,\Sigma_{\sss{NL}}(z-y).
\ee  
Or, in momentum representation:
\be
\label{top/self-energy}
  \tilde{\Sigma}(p,-p') 
  = 
  (2\pi)^4\,\delta^{(4)}(p-p')\,\Bigl[ \ps - m + \tilde{\Sigma}_{\sss{NL}}(p^2)
  \Bigr], 
\ee
where we performed the Fourier transforms
\begin{eqnarray}
  \Sigma_{\sss{NL}}(z-y) &=& \frac{1}{(2\pi)^4}\,\int d^4l\,e^{-il\cdot (z-y)}
                             \,\tilde{\Sigma}_{\sss{NL}}(l^2), 
                             \nonumber \\[1mm]
  \Sigma(y,z)            &=& \frac{1}{(2\pi)^8}\,\int d^4p\,d^4p'\,
                             e^{ip\cdot y}\,e^{-ip'\cdot z}\,
                             \tilde{\Sigma}(p,-p').
\end{eqnarray}
By convention we use a tilde to indicate functions in the momentum
representation. Upon inversion of Eq.~(\ref{top/self-energy}), the non-local 
coefficient is indeed seen to act as a mass-like correction to the free 
fermion propagator. 

Next we investigate how the gauge-boson--fermion--fermion vertex is modified 
by the non-local interaction. This vertex consists of two parts: a local and a
non-local one. The local part is standard:
\be
  \begin{picture}(40,35)(0,-2)
    \Photon(0,0)(40,0){2}{5}
    \ArrowLine(22,-12)(23,-12)
    \Line(12,-12)(22,-12)
    \ArrowLine(40,0)(60,30)
    \ArrowLine(60,-30)(40,0)
    \GCirc(40,0){3}{0}
    \Text(0,8)[lb]{$a,\mu$}
    \Text(0,-8)[lt]{$q$}
    \Text(60,20)[lt]{$p'$}
    \Text(60,-20)[lb]{$p$}
  \end{picture} 
  \hspace*{7ex} 
  :\ ig\,\Gamma_{\sss{L}}^{a,\,\mu}(x,y,z) 
  = 
  \frac{i\,\delta^3 {\cal S}_{\sss{L}}}{\delta A_{\mu}^a(x)\,\delta\psi(y)
                                        \,\delta\bar{\psi}(z)}
  \Biggl.\Biggr|_{A=\psi=\bar{\psi}=0} 
  =
  ig\,\T^a\,\delta^{(4)}(x-y)\,\delta^{(4)}(x-z)\,\gamma^{\mu}.
  \vspace*{2ex}
\ee
Or, in momentum representation:
\be
  ig\,\tilde{\Gamma}_{\sss{L}}^{a,\,\mu}(q,p,-p') 
  = 
  ig\,\T^a\,\gamma^{\mu}\,(2\pi)^4\,\delta^{(4)}(q+p-p'). 
\ee
In order to calculate the non-local contribution to the vertex, it is 
convenient to take the Fourier transform of $\Sigma_{\sss{NL}}$, perform a 
Taylor expansion of $\tilde{\Sigma}_{\sss{NL}}(l^2)$, and finally perform 
integration by parts. Then $\SNL$ can be rewritten as 
\begin{eqnarray}
\label{Snl_n}
  \SNL       &=& \sum_{n=0}^{\infty}\frac{1}{n!}\,{\cal S}_{\sss{NL}}^{(n)}\,
                 \biggl( \frac{d}{d l^2} \biggr)^n\,
                 \tilde{\Sigma}_{\sss{NL}}(l^2)\Biggl. \Biggr|_{l^2=0}
                 \equiv 
                 \sum_{n=0}^{\infty}\frac{1}{n!}\,{\cal S}_{\sss{NL}}^{(n)}\,
                 \tilde{\Sigma}_{\sss{NL}}^{(n)}(0),
                 \nonumber \\
  \SNL^{(n)} &=& \int d^4x\,d^4y\,\delta^{(4)}(x-y)\,\bar{\psi}(x)\,
                 (-\partial_y^2)^n\,\mbox{P}\exp\Biggl[ 
                 -\,ig\int\limits_x^y \TA_{\mu}(\omega)\,d\omega^{\mu} 
                 \biggr]\psi(y).
\end{eqnarray}
Of course, this expansion is not always allowed. We assume, however, that the 
non-local coefficient has analyticity properties that are similar to the ones 
expected for a normal self-energy function. As such we can perform the 
calculation in the regime where the above expansion is applicable and 
subsequently extend the range of validity by means of an analytical 
continuation. In fact, we will be able to present the non-local Feynman rules 
in such a way that the Ward identities are fulfilled irrespective of the 
precise form of the non-local coefficient.
The non-local part of the gauge-boson--fermion--fermion vertex now reads
\be
\label{nl/u1/4}
  ig\,\Gamma_{\sss{NL}}^{a,\,\mu}(x,y,z) 
  = 
  \frac{i\,\delta^3 \SNL}{\delta A_{\mu}^a(x)\,\delta\psi(y)\,
  \delta\bar{\psi}(z)}\Biggl.\Biggr|_{A=\psi=\bar{\psi}=0} 
  = 
  ig\,\T^a\,\sum_{n=0}^{\infty} \frac{1}{n!}\,
  \tilde{\Sigma}_{\sss{NL}}^{(n)}(0)\,\A^{\mu}_n(z,y|x),
\ee
where $\A_n^{\mu}(z,y|x)$ is given by
\be
  \A_n^{\mu}(z,y|x) 
  =
  -\,i\int d^4\tau\,\delta^{(4)}(z-\tau)\,(-\partial^2_{\tau})^n
  \,\delta^{(4)}(\tau-y)\int\limits_z^{\tau} \delta^{(4)}(\omega-x)\,
  d\omega^{\mu}.
\ee
The corresponding Fourier transform can be simplified by eliminating some 
$\delta$-function integrations and performing integration by parts:
\be
\label{nl/u1/5}
  \tilde{\A}_n^{\mu}(-p',p|q) 
  = 
  -\,i\int d^4\xi\,d^4\tau\,\delta^{(4)}(\xi-\tau)\,e^{ip'\cdot\xi}\, 
  (-\partial_{\tau}^2)^n\,e^{-ip\cdot\tau} 
  \int\limits_{\xi}^{\tau} e^{-iq\cdot\omega}\,d\omega^{\mu}.
\ee
By working out one of the $(-\partial_{\tau}^2)$ operators with the help of
Eq.~(\ref{U/der}), one can derive the recursion relation 
\be
  \tilde{\A}_n^{\mu}(-p',p|q) 
  = 
  p^2 \tilde{\A}_{n-1}^{\mu}(-p',p|q) 
  + (2p+q)^{\mu}\,(p+q)^{2n-2}\,(2\pi)^4\,\delta^{(4)}(q+p-p').
\ee
{}From the base of the recursion, $\tilde{\A}_0^{\mu}(-p',p|q)=0$, it is clear 
that all terms in the series will be proportional to 
$(2p+q)^{\mu}\,(2\pi)^4\,\delta^{(4)}(q+p-p')$. The solution of the recursion 
relation can be found in App.~\ref{app:a}:
\be
\label{A_n1}
  \tilde{\A}_n^{\mu}(-p',p|q) 
  = 
  (2p+q)^{\mu}\,\frac{(q+p)^{2n}-p^{2n}}{(q+p)^2-p^2}
  \,(2\pi)^4\,\delta^{(4)}(q+p-p').
\ee
Substituting this into the definition of the gauge-boson--fermion--fermion 
vertex (\ref{nl/u1/4}), one obtains for the non-local contribution in 
momentum representation
\begin{eqnarray}
  ig\,\tilde{\Gamma}_{\sss{NL}}^{a,\,\mu}(q,p,-p') &=& 
           ig\,\T^a\,(p+p')^{\mu}\sum_{n=0}^{\infty} \frac{1}{n!}\,
           \tilde{\Sigma}_{\sss{NL}}^{(n)}(0)\,
           \frac{p'^{\,2n}-p^{2n}}{p'^{\,2}-p^2}\,
           (2\pi)^4\,\delta^{(4)}(q+p-p')
           \nonumber \\[1mm]
                                                   &=&
           ig\,\T^a\,\frac{(p+p')^{\mu}}{p'^{\,2}-p^2}\, 
           \biggl[ \tilde{\Sigma}_{\sss{NL}}(p'^{\,2})
                   - \tilde{\Sigma}_{\sss{NL}}(p^2) \biggr]\,
           (2\pi)^4\,\delta^{(4)}(q+p-p').
\end{eqnarray}
This expression exhibits the proper infrared behaviour,
\be
  \tilde{\Gamma}_{\sss{NL}}^{a,\,\mu}(q,p,-p') 
  \stackrel{q\to 0}{\longrightarrow}
  \T^a\,(2\pi)^4\,\delta^{(4)}(q+p-p')\,
  \frac{\partial\,\tilde{\Sigma}_{\sss{NL}}(p^2)}{\partial p_{\mu}},
\ee
required for guaranteeing the usual eikonal factorization in the infrared 
limit. 

It is easy to verify that the so-obtained full gauge-boson--fermion--fermion 
vertex satisfies the Ward identity for dressed fermion propagators:
\be
  q_{\mu}\,\Bigl[ \tilde{\Gamma}_{\sss{L}}^{a,\,\mu}(q,p,-p') 
                 + \tilde{\Gamma}_{\sss{NL}}^{a,\,\mu}(q,p,-p') \Bigr]
  = 
  (2\pi)^4\,\delta^{(4)}(q+p-p')\,\T^a\,
  \Bigl[ \tilde{S}_F^{-1}(p')-\tilde{S}_F^{-1}(p) \Bigr],
\ee
with $\tilde{S}_F^{-1}(p) = \ps-m+\tilde{\Sigma}_{\sss{NL}}(p^2)$. From this 
we can conclude that the described non-local approach allows a gauge-invariant 
resummation of fermion self-energies, while at the same time reducing the 
complexity of the necessary gauge-restoring higher-point interactions to a 
minimum. The general higher-point interaction between two fermions and $k$ 
gauge bosons reads
\begin{eqnarray}
\label{2fkGB}
  ig^k\,\Gamma_{\sss{NL}}^{a_1..a_k,\,\mu_1..\mu_k} (x_1,..\,,x_k,y,z) 
  &=&  
  \frac{i\,\delta^{k+2} {\cal S}_{\sss{NL}}}
       {\delta A_{\mu_1}^{a_1}(x_1)\,..\,\delta A_{\mu_k}^{a_k}(x_k)\,
        \delta \psi(y)\,\delta\bar{\psi}(z)}
  \Biggl.\Biggr|_{A=\psi=\bar{\psi}=0}
  \nonumber \\[1mm]
  &=&
  ig^k \sum_{n=0}^{\infty} \frac{1}{n!}\,\tilde{\Sigma}_{\sss{NL}}^{(n)}(0)\,
  \sum\limits_{\sss{perm}}\A_n^{\nu_1..\nu_k}(z,y|y_1,..\,,y_k)\,
  \T^{b_1}\cdots\,\T^{b_k}.\hspace*{5ex}
\end{eqnarray}
The latter sum involves all possible permutations 
$\{(b_1,\nu_1,y_1),\ldots ,(b_k,\nu_k,y_k)\}$ of the basic set 
$\{(a_1,\mu_1,x_1),\ldots ,(a_k,\mu_k,x_k)\}$.
The Fourier transform of the path-ordered tensor-function  
$\A_n^{\mu_1..\mu_k}(z,y|x_1,..\,,x_k)$ is given by
\be
\label{A_nk}
  \tilde{\A}_n^{\mu_1..\mu_k}(-p',p|q_1,..\,,q_k)
  =
  (-i)^k\!\int d^4\xi\,d^4\tau\,\delta^{(4)}(\xi-\tau)\,e^{ip'\cdot\xi}
  (-\partial^2_{\tau})^n\,e^{-ip\cdot\tau}\,\prod\limits_{j=1}^k
  \,\,\int\limits_{\omega_{j\!-\!1}}^{\tau}d\omega_j^{\mu_j}\,
  e^{-iq_j\cdot\omega_j},
\ee
with $\omega_0 = \xi$. In App.~\ref{app:a} we discuss briefly the recursion 
relations corresponding to these tensor-functions and present the solution for 
general $k$. Based on the simple `Ward identities' for the tensor-functions 
given in App.~\ref{app:a}, it is easy to verify that the general interaction
(\ref{2fkGB}) obeys the Ward identity
\begin{eqnarray}
  \lefteqn{\hspace*{-2.0cm}
      q_{r,\,\mu_r}\,\tilde{\Gamma}_{\sss{NL}}^{a_1..a_k,\,\mu_1..\mu_k} 
      (q_1,..\,,q_k,p,-p') = \tilde{\Gamma}_{\sss{NL}}^{a_1..<a_r>..a_k,\,
      \mu_1..<\mu_r>..\mu_k}(q_1,..\,,<\!q_r\!>,..\,,q_k,p+q_r,-p')\,\T^{a_r}}
      \nonumber \\[4mm]
  & & -\,\T^{a_r}\,\tilde{\Gamma}_{\sss{NL}}^{a_1..<a_r>..a_k,\,
      \mu_1..<\mu_r>..\mu_k}(q_1,..\,,<\!q_r\!>,..\,,q_k,p,-p'+q_r) 
      \nonumber \\[2mm]
  & & -\,\sum\limits_{j\neq r} \,(\F^{a_r})^{a_j d}\,\biggl[
      \tilde{\Gamma}_{\sss{NL}}^{a_1..<a_r>..a_k,\,
      \mu_1..<\mu_r>..\mu_k}(q_1,..\,,<\!q_r\!>,..\,,q_k,p,-p')
      \biggl]_{\begin{array}{l}
                 \\[-3mm]
                 \!\!\scriptstyle{a_j \to d} \\[-1mm]
                 \!\!\scriptstyle{q_j \to q_j+q_r}
               \end{array}}
      \hspace*{-1cm}.
\end{eqnarray}
Here $<\!i_r\!>$ indicates that the index $i_r$ has been removed. 
The substitutions in the last term should be applied to the expression 
inside the brackets only. The $SU(N)$ generator in the adjoint representation 
$\F^{a_r}$ has been defined in the previous section.

{}From all this we can conclude that the above-described procedure allows the 
gauge-invariant resummation of fermion self-energies in the context of a 
$SU(N)$ symmetric theory. At the same time the compensating terms in the 
higher-point interactions are kept to a minimum. It should be
noted, however, that this procedure is not sufficient for a gauge-invariant 
description of unstable fermions in the Standard Model, since the symmetry is
explicitly broken in that case. We will come back to this point in 
Sect.~\ref{sec:sm}.


\section{An effective-Lagrangian approach for gauge bosons}
\label{sec:ww}

The next step is to extend the non-local method to the gauge-boson sector.
We remind the reader that the non-local Lagrangian should allow the Dyson 
resummation of the gauge-boson self-energies, in order to make the link to
unstable gauge bosons later on, and it should preserve gauge invariance with a 
minimum of additional higher-point interactions. The starting-point of the
non-local effective Lagrangian should therefore be a bilinear gauge-boson 
interaction. In the light of the discussion presented in Sect.~\ref{sec:fls}, 
the main idea is to rearrange the series on the right-hand side of 
Eq.~(\ref{fls/expansion}) in such a way that each term becomes gauge-invariant 
by itself. Subsequent truncation of the series at a given term is then 
allowed. 

Since the gauge bosons transform in the adjoint representation 
($\TF \to \TF'=\G \TF \G^{-1}$), the non-local action for gauge bosons differs
from the one for fermions in the way the path-ordered exponentials occur. 
For an $SU(N)$ Yang--Mills theory it takes the form
\be
\label{ww/action}
  \SNL 
  = 
  -\,\frac{1}{2} \int d^4x\,d^4y\,\Sigma_{\sss{NL}}(x-y)\, 
  \Tr\Bigl[ U(y,x)\,\TF_{\mu\nu}(x)\,U(x,y)\,\TF^{\mu\nu}(y) \Bigr],
\ee
or, expanding the non-local coefficient $\Sigma_{\sss{NL}}$ in terms of 
derivatives,
\begin{eqnarray}
  \SNL       &=& \sum\limits_{n=0}^{\infty} \frac{1}{n!}\, 
                 \tilde{\Sigma}_{\sss{NL}}^{(n)}(0)\,\SNL^{(n)},
                 \nonumber \\
  \SNL^{(n)} &=& -\,\frac{1}{2}\int d^4x\,d^4y\,\delta^{(4)}(x-y)\,
                 (-\partial_y^2)^n\,\Tr\Bigl[ U(y,x)\,\TF_{\mu\nu}(x)\,U(x,y)\,
                                              \TF^{\mu\nu}(y) \Bigr].
\end{eqnarray}
As required, the action contains bilinear gauge-boson interactions. The induced
infinite tower of higher-point gauge-boson interactions, which are also of
progressively higher order in the coupling constant $g$, is needed for 
restoring gauge invariance.

Let us derive the relevant Feynman rules, starting with the two-point 
function 
\be
  \begin{picture}(80,30)(0,-2)
    \Photon(0,0)(80,0){2}{10}
    \ArrowLine(25,-12)(26,-12)
    \Line(15,-12)(25,-12)
    \ArrowLine(55,-12)(54,-12)
    \Line(65,-12)(55,-12)
    \Text(0,8)[lb]{$a_1,\mu_1$}
    \Text(0,-8)[lt]{$q_1$}
    \Text(80,8)[rb]{$a_2,\mu_2$}
    \Text(80,-8)[rt]{$q_2$}
    \GCirc(40,0){3}{0}
  \end{picture} 
  \hspace*{5ex}
  :\
  i\,\Sigma^{a_1a_2,\,\mu_1\mu_2}(x_1,x_2) 
  = 
  \frac{i\,\delta^2 (\SL+\SNL)}{\delta A^{a_1}_{\mu_1}(x_1)\,
        \delta A^{a_2}_{\mu_2}(x_2)}\Biggl.\Biggr|_{A=0}, 
  \vspace*{2ex}
\ee
where the local action $\SL$ follows from the gauge-boson term in 
Eq.~(\ref{lagr}). The Fourier transform of this two-point function can be 
calculated in a straightforward way, since the path-ordered exponentials are
effectively unity. The result reads 
\be
\label{ww/self-energy}
  i\,\tilde{\Sigma}^{a_1a_2,\,\mu_1\mu_2}(q_1,q_2)
  =
  i\,\delta^{a_1a_2} \Bigl( q_1^{\mu}q_1^{\nu} - q_1^2 g^{\mu\nu} \Bigr)\,
  \Bigl[ 1 + \tilde{\Sigma}_{\sss{NL}}(q_1^2) \Bigr]
  \,(2\pi)^4\,\delta^{(4)}(q_1+q_2).
\ee
Note that this two-point interaction is transverse, as it should be for an 
unbroken theory. The non-local coefficient acts as a (dimensionless) 
correction to the transverse free gauge-boson propagator. So, exactly what is
needed for the Dyson resummation of the gauge-boson self-energies. 

The general non-local interaction between $k$ gauge bosons consists of four 
distinct contributions, with either two, three or four gauge fields supplied 
by the field strengths in Eq.~(\ref{ww/action}). In order to simplify the 
derivation of the Feynman rules, it is convenient to write 
Eq.~(\ref{ww/action}) in the adjoint representation rather than the 
fundamental representation:
\be
  \SNL 
  = 
  -\,\frac{1}{4} \int d^4x\,d^4y\,\Sigma_{\sss{NL}}(x-y)\, 
  F^a_{\mu\nu}(x)\,U^{ab}(x,y)\,F^{\mu\nu,\,b}(y), 
\ee
with
\be
  U^{ab}(x,y) 
  = 
  \mbox{P}\exp\Biggl[ -\,ig\int\limits_x^y \F^c A^c_{\mu}(\omega)\,
  d\omega^{\mu} \Biggr]^{ab}.
\ee
For the general interaction between $k$ gauge bosons we now obtain
\begin{eqnarray}
  \begin{picture}(75,45)(0,-2)
    \Photon(0,0)(40,0){2}{5}
    \ArrowLine(25,-12)(26,-12)
    \Line(15,-12)(25,-12)
    \Photon(40,0)(60,30){2}{5}
    \ArrowLine(57,8)(55,5)
    \Line(57,8)(62,15.5)
    \Photon(60,-30)(40,0){2}{5}
    \ArrowLine(57,-8)(55,-5)
    \Line(57,-8)(62,-15.5)
    \GCirc(40,0){3}{0}
    \Text(0,8)[lb]{$a_1,\mu_1$}
    \Text(0,-8)[lt]{$q_1$}
    \Text(60,35)[cb]{$a_2,\mu_2$}
    \Text(65,20)[lb]{$q_2$}
    \Text(60,-35)[ct]{$a_k,\mu_k$}
    \Text(65,-20)[lt]{$q_k$}
    \GCirc(75,12){0.5}{0}
    \GCirc(80,4){0.5}{0}
    \GCirc(80,-4){0.5}{0}
    \GCirc(75,-12){0.5}{0}
  \end{picture} 
  \hspace*{7ex} 
  &:& ig^{k-2}\,\Gamma_{\sss{NL}}^{a_1..a_k,\,\mu_1..\mu_k}(x_1,..\,,x_k) \ =\
      \frac{i\,\delta^k \SNL}{\delta A_{\mu_1}^{a_1}(x_1)\,..\,
                              \delta A_{\mu_k}^{a_k}(x_k)}
      \Biggl.\Biggr|_{A=0} \nonumber \\[4mm]
  & & =\  ig^{k-2} \sum_{n=0}^{\infty} \frac{1}{n!}\,
          \tilde{\Sigma}_{\sss{NL}}^{(n)}(0)\,\sum\limits_{\sss{perm}}
          (\F^{b_3}\cdots\,\F^{b_k})^{b_1b_2}\,
          V_{\sss{NL},\,n}^{\nu_1..\nu_k}(y_1,..\,,y_k),
\end{eqnarray}
where the sum involves all possible permutations
$\{(b_1,\nu_1,y_1),\ldots ,(b_k,\nu_k,y_k)\}$ of the basic set 
$\{(a_1,\mu_1,x_1),\ldots ,(a_k,\mu_k,x_k)\}$.
The Fourier transform of the path-ordered tensor-function 
$V_{\sss{NL},\,n}^{\mu_1..\mu_k}(x_1,..\,,x_k)$ can be expressed in terms of 
the path-ordered tensor-functions introduced in the previous section:   
\begin{eqnarray}
\label{ww/k-point}
  \tilde{V}_{\sss{NL},\,n}^{\mu_1..\mu_k}(q_1,..\,,q_k) 
  &=&
  \frac{1}{2}\,T^{\mu_1\mu_2}(q_1,q_2)\,
  \tilde{\A}_n^{\mu_3..\mu_k}(q_1,q_2|q_3,..\,,q_k) 
  \nonumber \\
  &+& 
  \frac{1}{4}\,A^{\mu_1,\,\mu_2\mu_k}(q_1)\,
  \tilde{\A}_n^{\mu_3..\mu_{k\!-\!1}}(q_1,q_2+q_k|q_3,..\,,q_{k\!-\!1})
  \nonumber \\
  &-&
  \frac{1}{4}\,A^{\mu_2,\,\mu_1\mu_3}(q_2)\,
  \tilde{\A}_n^{\mu_4..\mu_k}(q_1+q_3,q_2|q_4,..\,,q_k)
  \nonumber \\
  &-&
  \frac{1}{4}\,g^{\mu_1\mu_2}\,g^{\mu_3\mu_k}\,
  \tilde{\A}_n^{\mu_4..\mu_{k\!-\!1}}(q_1+q_3,q_2+q_k|q_4,..\,,q_{k\!-\!1}),
\end{eqnarray}  
where the first term contributes for $k\ge 2$, the second/third term for
$k\ge 3$, and the fourth term for $k\ge 4$.
Here we introduced the transverse tensors
\be
  T^{\mu\nu}(p,q)
  =
  (p \cdot q)\,g^{\mu\nu} - p^{\nu} q^{\mu},
  \qquad
  A^{\mu,\,\nu\rho}(q) 
  = 
  g^{\mu\nu}\,q^{\rho} - g^{\mu\rho}\,q^{\nu}.
\ee
These tensors have the following properties: $p^{\mu}\,T^{\mu\nu}(p,q)=
T^{\mu\nu}(p,q)\,q^{\nu} = q^{\mu}A^{\mu,\,\nu\rho}(q) = 0$,
$p^{\rho}A^{\mu,\,\nu\rho}(q) = T^{\mu\nu}(q,p)\,$ and 
$\,p^{\nu}A^{\mu,\,\nu\rho}(q) = - T^{\mu\rho}(q,p)$.
Using in addition the properties of the tensor-functions $\tilde{\A}_n$ given 
in App.~\ref{app:a}, one can verify that the general non-local gauge-boson
interaction satisfies the (ghost-free) Ward identity 
\begin{eqnarray}
  q_{r,\,\mu_r}\,\tilde{\Gamma}_{\sss{NL}}^{a_1..a_k,\,\mu_1..\mu_k}
  (q_1,..\,,q_k) 
  = 
  -\,\sum\limits_{j\neq r} \,(\F^{a_r})^{a_j d}\,\biggl[
  \tilde{\Gamma}_{\sss{NL}}^{a_1..<a_r>..a_k,\,\mu_1..<\mu_r>..\mu_k}
  (q_1,..\,,<\!q_r\!>,..\,,q_k) \biggl]_{\begin{array}{l}
                                           \\[-3mm]
                                           \!\!\scriptstyle{a_j \to d} \\[-1mm]
                                           \!\!\scriptstyle{q_j \to q_j+q_r}
                                         \end{array}}
  \hspace*{-1.0cm}. \hspace*{1.1ex}
\end{eqnarray}
On top of that, the non-local three-point interaction exhibits the proper
infrared behaviour,
\be
  \tilde{\Gamma}_{\sss{NL}}^{a_1a_2a_3,\,\mu_1\mu_2\mu_3}(q_1,q_2,q_3)
  \stackrel{q_1\to 0}{\longrightarrow}
  -\,(\F^{a_1})^{a_2a_3}\,(2\pi)^4\,\delta^{(4)}(q_1+q_2+q_3)\,
  \frac{\partial}{\partial q_{2,\,\mu_1}}
  \biggl\{ T^{\mu_2\mu_3}(q_2,-q_2)\,\tilde{\Sigma}_{\sss{NL}}(q_2^2) 
  \biggr\},
\ee
thereby guaranteeing the usual eikonal factorization in the infrared limit.

In addition to the non-local contributions, the three- and four-point 
gauge-boson interactions also receive contributions from the local action.
With our conventions these local contributions read 
\be
  ig^{k-2}\,\Gamma_{\sss{L}}^{a_1..a_k,\,\mu_1..\mu_k}(x_1,..\,,x_k)
  =
  ig^{k-2}\,\sum\limits_{\sss{perm}}(\F^{b_3}\cdots\,\F^{b_k})^{b_1b_2}\,
  V_{\sss{L}}^{\nu_1..\nu_k}(y_1,..\,,y_k),
\ee
with
\begin{eqnarray}
 \tilde{V}_{\sss{L}}^{\mu_1\mu_2\mu_3}(q_1,q_2,q_3)          &=&
        \frac{1}{2}\,A^{\mu_1,\,\mu_2\mu_3}(q_1)\,
        (2\pi)^4\,\delta^{(4)}(q_1+q_2+q_3),
        \nonumber \\[1mm]
 \tilde{V}_{\sss{L}}^{\mu_1\mu_2\mu_3\mu_4}(q_1,q_2,q_3,q_4) &=&
        -\,\frac{1}{4}\,g^{\mu_1\mu_2}g^{\mu_3\mu_4}\,
        (2\pi)^4\,\delta^{(4)}(q_1+q_2+q_3+q_4),
        \nonumber \\[2mm]
 \tilde{V}_{\sss{L}}^{\mu_1..\mu_k}(q_1,..\,,q_k)            &=& 
        0 \qquad (k>4).
\end{eqnarray}

Although the above-described non-local procedure provides a gauge-invariant 
framework for performing the Dyson resummation of the gauge-boson 
self-energies, we want to stress that it is not unique. We have seen in 
Sect.~\ref{sec:fls} that the FLS provides a different solution of the system 
of gauge-boson Ward identities. In the context of non-local effective 
Lagrangians it is always possible to add additional towers of gauge-boson 
interactions that start with three-point interactions and therefore do not 
influence the Dyson resummation of the gauge-boson self-energies. For instance,
the non-local action
\be
  \SNL' 
  =
  g\,\int d^4x\,d^4y\,d^4z\,V(x,y,z)\,\Tr\Bigl[ U(z,x)\,\TF^{\mu}_{\ \nu}(x)\, 
  U(x,y)\,\TF^{\nu}_{\ \rho}(y)\,U(y,z)\,\TF^{\rho}_{\ \mu}(z) \Bigr]
\ee
is gauge-invariant and does not affect the gauge-boson self-energies.
It does contribute, however, to the interaction between three or more gauge
bosons. As such it leads to a zero-mode solution of the system of 
gauge-boson Ward identities. In our quest for minimality we have opted to 
leave out such zero-mode solutions, as they are anyhow immaterial for the
discussion of self-energies. In the light of the discussion presented in 
Sect.~\ref{sec:fls}, we rearrange the series on the right-hand 
side of Eq.~(\ref{fls/expansion}) according to gauge-invariant towers of 
gauge-boson interactions labelled by the minimum number of gauge bosons that
are involved in the non-local interaction. Effectively this constitutes an 
expansion in powers of the coupling constant $g$, since a higher minimum
number of particles in the interaction is equivalent to a higher minimum 
order in $g$. In order to achieve minimality we have truncated this series at 
the lowest effective order.


\section{Unstable particles in the Standard Model}
\label{sec:sm}

In this section we address the case of phenomenological interest: unstable 
particles in a broken $[SU(3)_C\times ]SU(2)_L\times U(1)_Y$ gauge theory. 
First we briefly fix the notations. The $SU(2)_L\times U(1)_Y$ gauge-group 
element is defined as
\be
  G(x) = \exp\Bigl[ ig_2\T^a\theta^a(x) - ig_1\frac{Y}{2}\,\theta_Y(x) \Bigr],
\ee
where the $SU(2)$ generators $\T^a$ can be expressed in terms of the standard 
Pauli spin matrices $\sigma^a$ $\,(a=1,2,3)$ according to $\T^a=\sigma^a/2$. 
The normalization condition and commutation relation read 
$\Tr(\sigma^a\sigma^b) = 2\delta^{ab}$ and 
$\Bigl[ \sigma^a,\sigma^b \Bigr] = 2i\epsilon^{abc}\,\sigma^c$, with the
$SU(2)$ structure constant $\epsilon^{abc}$ given by
\be
  \epsilon^{abc} 
  =
  \left\{ \begin{array}{cl}
            +1 & \mbox{\ if }(a,b,c)=\mbox{\ even permutation of }(1,2,3) \\
            -1 & \mbox{\ if }(a,b,c)=\mbox{\ odd permutation of }(1,2,3)\ \ .\\
            0  & \mbox{\ else}
          \end{array}
  \right.
\ee
The $SU(2)$ generators in the adjoint representation are given by 
$(\F^a)^{bc} = -i\epsilon^{abc}$.

\subsection{The gauge bosons}
\label{sec:sm/wz}

In the Standard Model there are four gauge fields, $W^a_{\mu}$ $\,(a=1,2,3)$
and $B_{\mu}$, with the corresponding field-strength tensors given by
\be
  \TF_{\mu\nu} = \partial_\mu \TW_\nu - \partial_\nu \TW_\mu 
                 - i g_2\,\bigl[ \TW_\mu,\TW_\nu \Bigr],
  \qquad 
  B_{\mu\nu} = \partial_\mu B_\nu - \partial_\nu B_\mu,
\ee
using the shorthand notations
\be
  \TF_{\mu\nu} \equiv \T^a F_{\mu\nu}^a,
  \qquad
  \TW_{\mu} \equiv \T^a W_{\mu}^a.
\ee
In this notation the Yang--Mills Lagrangian reads:
\be
\label{sm/yang-mills}
  \LL^{\sss{YM}}(x) 
  = 
  -\,\frac{1}{2}\,\Tr\Bigl[ \TF_{\mu\nu}(x)\,\TF^{\mu\nu}(x) \Bigr]
  -\frac{1}{4}\,B_{\mu\nu}(x)\,B^{\mu\nu}(x).
\ee
The physically observable gauge-boson states are given by
\be
  W^{\pm}_{\mu} = \frac{1}{\sqrt{2}} (W^1_{\mu}\mp i\,W^2_{\mu}),
  \quad
  Z_{\mu} = c_W W^3_{\mu} + s_W B_{\mu},
  \quad 
  A_{\mu} = c_W B_{\mu} - s_W W^3_{\mu},
\ee
where $\,c_W=g_2/\sqrt{g_1^2+g_2^2}\,$ and $\,s_W=\sqrt{1-c_W^2}\,$ are the 
cosine and sine of the weak mixing angle. The electric charge is given by 
$e=g_1 g_2/\sqrt{g_1^2+g_2^2}$.

For the gauge-invariant treatment of unstable gauge bosons we can use a 
non-local Lagrangian that generates the relevant self-energy effects. 
The corresponding action can be split into two pieces. One piece is already 
known from the unbroken theory, bearing in mind that we have two 
field-strength tensors to work with:
\begin{eqnarray}
\label{sm/action}
  \SNL^{\sss{YM}} &=& -\,\frac{1}{4} \int d^4x\,d^4y\,\Sigma_1(x-y)\,
           B_{\mu\nu}(x)\,B^{\mu\nu}(y) \nonumber \\[1mm]
                  & & -\,\frac{1}{2} \int d^4x\,d^4y\,\Sigma_2(x-y)\,\Tr\Bigl[ 
           U_2(y,x)\,\TF_{\mu\nu}(x)\,U_2(x,y)\,\TF^{\mu\nu}(y) \Bigr].
\end{eqnarray}
Note that the path-ordered exponentials vanish in the first term. They are also
not needed, since $B_{\mu\nu}(x)$ is gauge-invariant by itself. In the second
term $U_2$ is the path-ordered exponential corresponding to $SU(2)_L$ 
[defined according to Eq.~(\ref{Pexp})]. Furthermore,
it is impossible to construct a gauge-invariant non-local operator 
of the form $B\cdot F$ using only gauge fields. Such interactions require
some additional fields with non-zero vacuum expectation value, i.e.~the 
Higgs fields. For the second piece of the non-local Lagrangian we therefore 
exploit the fact that the theory is spontaneously broken:
\begin{eqnarray}
\label{sm/action/higgs}
  \SNL^{\Phi} &=& -\,\frac{g_1g_2}{2M_W^2} \int d^4x\,d^4y\,\Sigma_3(x-y)\,
                  [ \Phi^{\dagger}(x)\,\TF_{\mu\nu}(x)\,\Phi(x) ]\,
                  B^{\mu\nu}(y) \nonumber \\
              & & -\,\frac{g_2^4}{4M_W^4} \int d^4x\,d^4y\,\Sigma_4(x-y)\,
                  [ \Phi^{\dagger}(x)\,\TF_{\mu\nu}(x)\,\Phi(x) ]\,
                  [ \Phi^{\dagger}(y)\,\TF^{\mu\nu}(y)\,\Phi(y) ],
\end{eqnarray}
with the ($Y\!=\!1$) Higgs doublet $\Phi$ given by
\be
\label{higgs}
  \Phi(x)
  =                     
  \left( \begin{array}{c}
            \phi^+(x) \\[1mm]
            [v+H(x)+i\,\chi(x)]/\sqrt{2}
         \end{array} 
  \right)\ .
\ee
The non-zero vacuum expectation value $v$ is given by $v=2M_W/g_2$. 
The field operators contained in this additional effective action are clearly 
of higher dimension than the ones contained in the previously encountered
effective actions (see the prefactors $1/M_W^2$ and $1/M_W^4$). As such these
higher-dimensional operators have no local analogue in the Standard Model 
Lagrangian. They are required for achieving an explicit breaking of the 
$SU(2)$ symmetry amongst the $SU(2)$ gauge bosons in the transverse sector. 
After all, the loop effects in the Standard Model also lead to such explicit 
symmetry-breaking effects.   

For completeness we now list the two-point gauge-boson interactions induced by 
the above-specified non-local operators:
\be
  \begin{picture}(80,30)(0,-2)
    \Photon(0,0)(80,0){2}{10}
    \ArrowLine(25,-12)(26,-12)
    \Line(15,-12)(25,-12)
    \ArrowLine(55,-12)(54,-12)
    \Line(65,-12)(55,-12)
    \Text(0,8)[lb]{$V_1,\mu_1$}
    \Text(0,-8)[lt]{$q_1$}
    \Text(80,8)[rb]{$V_2,\mu_2$}
    \Text(80,-8)[rt]{$q_2$}
    \GCirc(40,0){3}{0}
  \end{picture} 
  \hspace*{5ex}
  :\
  i\,\tilde{\Sigma}^{\mu\nu}_{\sss{NL}}(q_1,q_2)
  =
  i\,\Bigl( q_1^{\mu} q_1^{\nu} - q_1^2 g^{\mu\nu} \Bigr)\,
  \tilde{\Pi}_{\sss{NL}}^{V_1V_2}(q_1^2)\,(2\pi)^4\,\delta^{(4)}(q_1+q_2),
  \vspace*{2ex}
\ee
with the transverse (dimensionless) self-energies given by
\begin{eqnarray}
\label{sm/self-energies}
  \tilde{\Pi}_{\sss{NL}}^{WW}(q_1^2)           \!&=&\! 
        \tilde{\Sigma}_2(q_1^2) \nonumber \\[1mm]
  \tilde{\Pi}_{\sss{NL}}^{ZZ}(q_1^2)           \!&=&\! 
        s_W^2\,\tilde{\Sigma}_1(q_1^2) + c_W^2\,\tilde{\Sigma}_2(q_1^2) 
        - 2s_W^2\,\tilde{\Sigma}_3(q_1^2) + c_W^2\,\tilde{\Sigma}_4(q_1^2) 
        \nonumber \\[1mm]
  \tilde{\Pi}_{\sss{NL}}^{\gamma Z}(q_1^2)     \!&=&\! 
        \tilde{\Pi}_{\sss{NL}}^{Z\gamma}(q_1^2) \ =\ 
        s_W c_W\,\tilde{\Sigma}_1(q_1^2) - s_W c_W\,\tilde{\Sigma}_2(q_1^2) 
        + (s_W^2-c_W^2)\,\frac{s_W}{c_W}\,\tilde{\Sigma}_3(q_1^2) 
        - s_W c_W\,\tilde{\Sigma}_4(q_1^2) \nonumber \\
  \tilde{\Pi}_{\sss{NL}}^{\gamma\gamma}(q_1^2) \!&=&\! 
        c_W^2\,\tilde{\Sigma}_1(q_1^2) + s_W^2\,\tilde{\Sigma}_2(q_1^2)
        + 2s_W^2\,\tilde{\Sigma}_3(q_1^2) + s_W^2\,\tilde{\Sigma}_4(q_1^2).
\end{eqnarray}
Thus four self-energies are parametrized by four independent functions. As such
all mass effects can be taken into account properly. If the theory would have
been unbroken, only two functions ($\Sigma_{1,2}$) would be available for 
parametrizing the four self-energies in the massless limit and two relations
among the self-energies would emerge. These relations hold indeed in the FLS 
if all fermions are massless (including the top-quark)~\cite{fls2}. 
At high energies $\Sigma_3$ and $\Sigma_4$ should vanish, since effectively 
the theory becomes unbroken.

At this point we remind the reader that we have only considered non-local
contributions to the transverse gauge-boson self-energies, which can be
resummed into dressed transverse gauge-boson propagators. In principle one 
should add also non-local terms that contribute to the longitudinal gauge-boson
self-energies 
($\tilde{\Sigma}_{\sss{long}}^{\mu\nu} \propto q_1^{\mu}q_1^{\nu}$), which
can be resummed into dressed longitudinal gauge-boson propagators. Since the
resummation in the transverse and longitudinal sectors can be performed 
independently, the longitudinal sector with its close relation to the 
gauge-fixing procedure can be treated separately. In view of minimality we 
refrain from adding non-local longitudinal terms. In physical matrix elements
the longitudinal propagators do not generate resonances and therefore there 
is no strict need for resumming (imaginary parts of) longitudinal 
self-energies. The imaginary parts that appear in the resummed transverse 
propagators of the $W,Z$ bosons are directly linked to the corresponding
decay widths $\Gamma_{W,Z}$ and are hence sufficient for a proper description 
of the resonance effects. In the covariant $R_{\xi}$ gauge 
\begin{eqnarray}
  \LL_{R_{\xi}}^{\sss{gauge fix}}(x) &=& - \,\frac{1}{2}\,\Biggl\{
          \frac{1}{\xi_{\gamma}}\,\Bigl[\partial^{\mu} A_{\mu}(x)\Bigr]^{2}
          + \frac{1}{\xi_{Z}}\,\Bigl[\partial^{\mu} Z_{\mu}(x) 
                                     - \xi_{Z}M_{Z}\chi(x)\Bigr]^{2} 
          \nonumber \\
                                     &+& \frac{2}{\xi_{W}}\, 
          \Bigl[\partial^{\mu} W_{\mu}^{+}(x) - i \xi_{W}M_{W}\phi^{+}(x)\Bigr]
          \Bigl[\partial^{\nu} W_{\nu}^{-}(x) + i \xi_{W}M_{W}\phi^{-}(x)\Bigr]
                                                          \Biggr\},
\end{eqnarray}
for instance, we 
obtain the following dressed gauge-boson propagators ($V=\gamma,Z,W$):
\begin{eqnarray}
\label{prop/W}
  P^{VV}_{\mu\nu}(q,\xi_{_V}) &=& -i\,D_T^{VV}(q^2)\,
            \biggl( g_{\mu\nu} - \frac{q_{\mu}q_{\nu}}{q^2} \biggr)
            - \frac{i\xi_{_V}}{q^2-\xi_{_V}M_V^2}\,\frac{q_{\mu}q_{\nu}}{q^2},
            \nonumber \\[1mm]
  P^{\gamma Z}_{\mu\nu}(q)    &=& P^{Z\gamma}_{\mu\nu}(q)
            \ =\ -i\,D_T^{\gamma Z}(q^2)\,
            \biggl( g_{\mu\nu} - \frac{q_{\mu}q_{\nu}}{q^2} \biggr). 
\end{eqnarray}
The transverse propagator functions $D_T$ are given by
\begin{eqnarray}
  D_T^{WW}(q^2)           &=& 
     \Bigl\{ q^2 - M_W^2 + q^2\,\tilde{\Pi}_{\sss{NL}}^{WW}(q^2) \Bigr\}^{-1}, 
     \nonumber \\[1mm]
  D_T^{\gamma\gamma}(q^2) &=&
     \Bigl[ q^2 - M_Z^2 + q^2\,\tilde{\Pi}_{\sss{NL}}^{ZZ}(q^2) \Bigr]/D(q^2),
     \nonumber \\[1mm]
  D_T^{ZZ}(q^2)           &=& 
     \Bigl[ q^2+q^2\,\tilde{\Pi}_{\sss{NL}}^{\gamma\gamma}(q^2) \Bigr]/D(q^2),
     \nonumber \\[1mm]
  D_T^{\gamma Z}(q^2)     &=& 
     - q^2\,\tilde{\Pi}_{\sss{NL}}^{\gamma Z}(q^2)/D(q^2),
     \nonumber \\[1mm]
  D(q^2)                  &=&
     \Bigl[ q^2 - M_Z^2 + q^2\,\tilde{\Pi}_{\sss{NL}}^{ZZ}(q^2) \Bigr]\,
     \Bigl[ q^2 + q^2\,\tilde{\Pi}_{\sss{NL}}^{\gamma\gamma}(q^2) \Bigr]
     - \Bigl[ q^2\,\tilde{\Pi}_{\sss{NL}}^{\gamma Z}(q^2) \Bigr]^2,
\end{eqnarray}
with the explicit mass terms originating from the Higgs part of the Standard
Model Lagrangian [see Eq.~(\ref{SMHiggs}) below].
However, this is not the complete story. We will have to redefine the 
photon field and the electromagnetic coupling, which are by definition 
identified by means of the $ee\gamma$ interaction in the Thomson limit 
($q_\gamma^2=0$). Since the $ee\gamma$ vertex does not receive non-local
contributions, only the non-local photonic self-energy contributions have to 
be adjusted. This results in a `finite renormalization' of the form  
\begin{eqnarray}
  && \left. q^2\,D_T^{\gamma\gamma}(q^2) \right|_{q^2=0} =
     \frac{1}{1+\tilde{\Pi}_{\sss{NL}}^{\gamma\gamma}(0)}  \to 1,
     \nonumber \\[1mm]
  && \left. q^2\,\tilde{\Pi}_{\sss{NL}}^{\gamma Z}(q^2) \right|_{q^2=0} \to 0.
\end{eqnarray} 
  
In App.~\ref{app:b} we list the Feynman rules for the relevant non-local 
three- and four-point interactions, needed for a gauge-invariant treatment of
reactions like $2f \to 4f,\,4f\gamma,\,6f$.

\subsection{The Higgs boson}
\label{sec:sm/higgs}

In the Standard Model the Higgs part of the Lagrangian is given by
\be
\label{SMHiggs}
  \LL^H(x) 
  =
  \biggl( D_{\mu}\Phi(x) \biggr)^{\dagger} \biggl( D^{\mu}\Phi(x) \biggr)
  + \mu^{2}\,\Bigl[ \Phi^{\dagger}(x)\,\Phi(x) \Bigr]   
  - \frac{\lambda}{4}\,\Bigl[ \Phi^{\dagger}(x)\,\Phi(x) \Bigr]^2,
\ee 
with the covariant derivative defined as
\be
  D_{\mu} 
  = 
  \partial_{\mu} - i g_2 \T^a W^a_{\mu} + i g_1 \frac{Y}{2}\,B_{\mu}.
\ee
The ($Y\!=\!1$) complex Higgs doublet $\Phi$ is expanded around its vacuum 
expectation value 
$\,<\!|\phi|\!>_0 = v/\sqrt{2} = \sqrt{2}\,\mu/\sqrt{\lambda}\,$ 
according to Eq.~(\ref{higgs}). The resulting Lagrangian describes one physical
scalar particle $H$ with mass $\,M_H = \sqrt{2}\,\mu\,$ and three degrees 
of freedom that are absorbed by the gauge bosons and that are hence rendered 
unphysical. Our aim in this subsection is to construct an effective Lagrangian
that generates a self-energy for the physical Higgs boson. At the same time we 
want to avoid generating any self-energies for the unphysical Higgs modes or 
the corresponding longitudinal gauge-boson modes. This is based on the same 
philosophy as adopted in the previous subsection. In order to achieve this 
aim we are led to a construction with only singlets, i.e.~without path-ordered 
exponentials: 
\be
\label{sm/NL/higgs}
  \SNL^H
  = 
  \frac{1}{2 v^2}\,\int d^4x\,d^4y\,\Sigma_H(x-y)\,
  \Biggl(\Bigl|\Phi(x)\Bigr|^{2}-\frac{v^2}{2}\Biggr)\,
  \Biggl(\Bigl|\Phi(y)\Bigr|^{2}-\frac{v^2}{2}\Biggr).
\ee
This Lagrangian induces the required self-energy in the physical Higgs 
propagator, without generating additional self-energy or tadpole contributions.
The combined local and non-local contributions to the two-point interaction
between physical Higgs bosons read:
\be
  \begin{picture}(80,30)(0,-2)
    \DashLine(0,0)(80,0){8}
    \ArrowLine(25,-12)(26,-12)
    \Line(15,-12)(25,-12)
    \ArrowLine(55,-12)(54,-12)
    \Line(65,-12)(55,-12)
    \Text(0,8)[lb]{$H$}
    \Text(0,-8)[lt]{$q_1$}
    \Text(80,8)[rb]{$H$}
    \Text(80,-8)[rt]{$q_2$}
    \GCirc(40,0){3}{0}
  \end{picture} 
  \hspace*{5ex}
  :\
  i\,\tilde{\Sigma}(q_1,q_2)
  =
  i\,\Bigl[ q_1^2-M_H^2+\tilde{\Sigma}_H(q_1^2) \Bigr]\,
  (2\pi)^4\,\delta^{(4)}(q_1+q_2),
  \vspace*{2ex}
\ee
which can be inverted trivially to give the dressed Higgs-boson propagator
\be 
\label{prop/H}
  P^H(q) = \frac{i}{q^2-M_H^2+\tilde{\Sigma}_H(q^2)}.
\ee
The remaining non-local scalar interactions can be found in App.~\ref{app:b}. 

This time we can explicitly check the proposed gauge-invariant resummation 
procedure by considering reactions like $\phi^+\phi^- \to \chi\chi$ in the 
limit $M_H^2 \gg q^2 \gg M_{W,Z}^2$. Indeed, we find that in leading 
approximation the vertex and box graphs are identical to the non-local three- 
and four-point interactions, provided that tadpole renormalization is applied.
This is caused by the fact that the scalar three- and four-point functions 
reduce to two-point functions as a result of the exchange of the heavy 
(physical) Higgs bosons. It should be noted that the check only works for the
leading terms, since the sub-leading contributions will already contain 
information on higher-order non-local towers (e.g.~the ones that start at 
three-point level).

\subsection{The top-quark}
\label{sec:sm/top}

In the Standard Model the fermions acquire their mass through the Yukawa
interactions with the Higgs field. Since all masses are basically different,
the $SU(2)$ symmetry is explicitly broken in the fermionic doublets. As such
the method described in Sect.~\ref{sec:top} is not applicable to the 
resummation of fermion self-energies in the Standard Model. In this subsection 
we concentrate on the top-quark, which has a large decay width and therefore 
can be described by means of perturbative methods.
We start off by writing down the Yukawa interaction for the top-quark:
\be
\label{sm/top/yukawa}
  \LL_{\sss{Yu}}^t(x) 
  = 
  -\,f_t\,\overline{Q}_L(x)\,\tilde{\Phi}(x)\,t_R(x)  + \mbox{h.c.},
\ee
where $f_t$ is the top-quark Yukawa coupling. The doublet $\overline{Q}_L(x)$
and singlet $t_R(x)$ can be expressed in terms of the top-quark and 
bottom-quark fields $t(x),b(x)$ according to
\be 
  Q_L(x) = \left( \begin{array}{c}
                    t_L(x) \\[1mm]
                    b_L(x)
                  \end{array} \right)
\ee
with
\be
  t_L(x) = \frac{1-\gamma_5}{2}\,t(x),
  \qquad
  t_R(x) = \frac{1+\gamma_5}{2}\,t(x)
  \quad \mbox{and} \quad
  b_L(x) = \frac{1-\gamma_5}{2}\,b(x).
\ee
In order to give a mass to the top-quark, a Higgs doublet $\tilde{\Phi}$ with 
opposite hypercharge ($Y\!=\!-1$) is required: 
\be 
  \tilde{\Phi}(x)
  =
  \left( \begin{array}{c}
           [v+H(x)-i\,\chi(x)]/\sqrt{2}\\[1mm]
           -\,\phi^-(x) 
         \end{array} \right)
  \quad \mbox{with} \quad
  \phi^-(x) = \Bigl[ \phi^+(x) \Bigr]^{\dagger}.
\ee
The resulting top-quark mass is given by $m_t=vf_t/\sqrt{2}$. Based on this
Yukawa interaction it is not difficult to construct a non-local effective 
action that generates a mass-like top-quark self-energy. There are two ways to 
non-localize the three fields in Eq.~(\ref{sm/top/yukawa}).
The first one involves a non-local interaction between $SU(2)$ singlets:
\be
\label{sm/NL/top}
  \SNL^t 
  = 
  \frac{\sqrt{2}}{v}\,\int d^4x\,d^4y\,\Sigma_t(x-y)\,\biggl\{
  \Bigl[\, \overline{Q}_L(x)\,\tilde{\Phi}(x) \,\Bigr]
  \,U_1(x,y)\,U_3(x,y)\,t_R(y) 
  + \mbox{h.c.} \biggr\},
\ee
where $U_1$ and $U_3$ are the path-ordered exponentials corresponding
to $U(1)_Y$ (for $Y=4/3$) and $SU(3)_C$, respectively. The latter path-ordered
exponential enters as a result of the fact that the top-quark also carries a 
colour charge. The second way of non-localizing Eq.~(\ref{sm/top/yukawa})
involves a non-local interaction between the $SU(2)$ doublets 
$\overline{Q}_L(x)$ and $[\tilde{\Phi}(y)t_R(y)]$, connected by a string of 
path-ordered exponentials $U_1(x,y)\,U_2(x,y)\,U_3(x,y)$. Note, that both 
effective Lagrangians contribute to the top-quark two-point interaction in the 
same way and therefore both allow a gauge-invariant resummation of the 
self-energy. A particular choice can be made on the basis of either explicit 
physical requirements (like the properties under parity transformations) or 
minimality considerations. In this paper we consider in detail the simplest of 
the two effective Lagrangians, given by Eq.~(\ref{sm/NL/top}).

The combined local and non-local contributions to the top-quark 
two-point interaction then read:
\be 
  \begin{picture}(60,15)(0,-2)
    \ArrowLine(0,0)(30,0)
    \ArrowLine(30,0)(60,0)
    \Text(0,8)[lb]{$t$}
    \Text(0,-8)[lt]{$p$}
    \Text(60,8)[rb]{$t$}
    \Text(60,-5)[rt]{$p'$}
    \GCirc(30,0){3}{0}
  \end{picture}
  \hspace*{5ex}
  :\ 
  i\,\tilde{\Sigma}(p,-p') 
  = 
  i\,(2\pi)^4\,\delta^{(4)}(p-p')\,
  \Bigl[ \ps - m_t + \tilde{\Sigma}_t(p^2) \Bigr],
  \vspace*{2ex} 
\ee
which results in the following dressed top-quark propagator:
\be 
\label{prop/t}
  P^t(p) = \frac{i}{\ps - m_t + \tilde{\Sigma}_t(p^2)}.
\ee

At first sight the effective action (\ref{sm/NL/top}) seems to have little in
common with the effective action introduced in Sect.~\ref{sec:top}. However,
the part originating from the non-zero vacuum expectation value of the Higgs,
which hence only involves fermions and gauge bosons, has the familiar form
\be
\label{sm/NL/top/nohiggs}
  \SNL^{t,\,v} 
  = 
  \int d^4x\,d^4y\,\Sigma_t(x-y)\,\bar{t}(x)\,U_1(x,y)\,U_3(x,y)\,t(y).
\ee
In App.~\ref{app:b} we list the Feynman rules for the various non-local 
three- and four-point interactions. Particularly noteworthy are the mixed
QCD--electroweak interactions, involving both gluons and electroweak bosons,
which are needed for the construction of gauge-invariant resummed amplitudes 
in certain mixed QCD--electroweak processes (like $e^{+}e^{-}\to t\bar{t}g$).

\subsection{Some simple examples}
\label{sec:sm/examples}

A substantial simplification occurs when all non-local coefficients are taken 
to be delta-functions:
\be
  \Sigma_j(x-y) = \Sigma_j\,\delta(x-y) \qquad (j=1,2,3,4,H,t).
\ee
By choosing appropriate values for the complex constants $\Sigma_j$, the 
simplified set of non-local actions can be used to implement the decay widths 
of unstable particles in a concise, gauge-invariant way. Strictly speaking, 
however, the proposed simplification is not supported by the actual loop 
effects in gauge theories, where no imaginary parts occur for space-like 
momenta. Nevertheless, it has become a very popular (ad hoc) procedure.

According to Eqs.~(\ref{prop/H}) and (\ref{prop/t}), the simplification 
correponds to constant shifts in the Higgs and top-quark propagators:
\be
 P^H(q) = \frac{i}{q^2 - M_H^2 + \Sigma_H}
 \qquad \mbox{and} \qquad
 P^t(p) = \frac{i}{\ps - m_t + \Sigma_t}.
\ee
As a result of the delta-functions in the non-local coefficients, the effective
Lagrangians defined by $\SNL^H$ in Eq.~(\ref{sm/NL/higgs}) and $\SNL^t$ in 
Eq.~(\ref{sm/NL/top}) become proportional to the corresponding local 
Lagrangians:
\be
  \LL_{\sss{NL}}^H(x) = -\,\Sigma_H\,\frac{\LL_{\sss{pot}}^H(x)}{M_H^2}
  \qquad \mbox{and} \qquad
  \LL_{\sss{NL}}^t(x) = -\,\Sigma_t\,\frac{\LL_{\sss{Yu}}^t(x)}{m_t},
\ee
with $\LL_{\sss{pot}}^H$ indicating the part of the local Higgs Lagrangian
that corresponds to the Higgs potential. From this it is clear that the 
combined effect of all non-local interactions amounts to the mere effective
replacements 
\be
M_H^2 \to M_H^2 - \Sigma_H 
\qquad \mbox{and} \qquad  
m_t \to m_t - \Sigma_t
\ee
in the Standard Model Lagrangian. Note that for imaginary shifts 
(e.g.~$\Sigma_H=iM_H\Gamma_H$) this procedure resembles the so-called 
fixed-width scheme.%
\footnote{By appropriately choosing the complex constants $\Sigma_{t,H}$, it is
          also possible to effectively replace the masses by the complex poles
          of the propagators [e.g.~with 
          $\Sigma_H = iM_H^2\Gamma_H/(M_H+i\,\Gamma_H)$ one obtains the 
          effective replacement $M_H^2 \to M_H^2/(1+i\,\Gamma_H/M_H)$]. Such a 
          complex pole mass is often better suited for the description of 
          resonances associated with heavy unstable particles~\cite{massdef}.}
However, in contrast to our non-local approach, the fixed-width scheme applies 
the effective replacements only to the propagators. So, in general the 
fixed-width scheme has to be adapted whenever the mass terms in the 
higher-point interactions play a role.

In the gauge-boson sector we need a further simplification. 
The higher-dimensional operators in $\SNL^{\Phi}$ have no Standard Model
analogues. Therefore, a large number of compensating higher-point interactions
remain, even if all non-local coefficients are taken to be delta-functions.
At this point we can exploit the fact that we don't strictly need all four
non-local gauge-boson coefficients for a gauge-invariant treatment of unstable
$W$ and $Z$ bosons. In order to properly generate the two corresponding decay
widths, it is formally sufficient to have only two independent non-local
coefficients. A huge simplification is achieved by setting 
$\Sigma_{3,4}=0$. This comes at a price, though. As mentioned in 
Sect.~\ref{sec:sm/wz}, two relations among the gauge-boson 
self-energies will emerge, like in the unbroken theory. This means that the 
self-energies involving photons are not independent anymore. In fact this is
not a real problem, since we will anyhow have to redefine the photon field and 
the electromagnetic coupling according to the $ee\gamma$ interaction in the
Thomson limit ($q_\gamma^2=0$).

The net effect of the simplifications in the gauge-boson sector amounts to a
rescaling of the $U(1)_Y$ and $SU(2)_L$ terms in the Yang--Mills Lagrangian 
(\ref{sm/yang-mills}):
\be
\label{examples/YM}
  \LL^{\sss{YM}}(x) + \LL_{\sss{NL}}^{\sss{YM}}(x) 
  = 
  -\,\frac{1}{2}\,(1+\Sigma_2)\,
  \Tr\Bigl[ \TF_{\mu\nu}(x)\,\TF^{\mu\nu}(x) \Bigr]
  -\frac{1}{4}\,(1+\Sigma_1)\,B_{\mu\nu}(x)\,B^{\mu\nu}(x). 
\ee
These rescaling factors can be absorbed into the gauge-boson fields and the
coupling constants according to
\be
\label{redef} 
 W^a = \frac{{W'}^a}{\sqrt{1+\Sigma_2}},
 \quad 
 g_2 = g'_2\,\sqrt{1+\Sigma_2}
 \qquad \mbox{and} \qquad
 B   = \frac{B'}{\sqrt{1+\Sigma_1}},
 \quad 
 g_1 = g'_1\,\sqrt{1+\Sigma_1}.
\ee
In terms of the redefined fields and couplings the Lagrangian 
(\ref{examples/YM}) retrieves the original Yang--Mills form. At the same time
the other Standard Model interactions are not changed by the redefinitions, as
the covariant derivatives stay the same. So, the only noticeable 
changes involve the gauge-boson mass matrix and consequently the $W^3$--$B$ 
mixing, which are both defined in terms of the coupling constants:
\begin{eqnarray}
\label{examples/YM/upshot}
  M_W^2 = \frac{1}{4}\,v^2 g_2^2         &\to & 
  {M'_W\!\!}^2 = \frac{M_W^2}{1+\Sigma_2},
  \nonumber \\[1mm]
  M_Z^2 = \frac{1}{4}\,v^2 (g_1^2+g_2^2) &\to &
  {M'_Z\!}^2 = M_Z^2\,\frac{1+c_w^2\Sigma_1+s_w^2\Sigma_2}
                         {(1+\Sigma_1)(1+\Sigma_2)} 
           \equiv \frac{M_Z^2}{1+C_Z},
  \nonumber \\[1mm]
  c_w = M_W/M_Z                          &\to & 
  c'_w = M'_W/M'_Z = c_w\,\sqrt{\frac{1+C_Z}{1+\Sigma_2}}.
\end{eqnarray}
For an imaginary non-local coefficient $\Sigma_2 = i\,\Gamma_W/M_W$ the 
redefined $W$ mass is identical to the so-called complex pole mass 
${M'_W\!\!}^2 = (M_W^2-iM_W\Gamma_W)/(1+\Gamma_W^2/M_W^2)$. A similar pole mass
can be obtained for the $Z$ boson by choosing $\Sigma_1$ in such a way that 
$C_Z = i\,\Gamma_Z/M_Z$.%
\footnote{By choosing $\Sigma_2 = i\,\Gamma_W/(M_W-i\,\Gamma_W)$ and 
          $C_Z = i\,\Gamma_Z/(M_Z-i\,\Gamma_Z)$ one can obtain the usual 
          fixed-width masses ${M'_V\!}^2 = M_V^2-iM_V\Gamma_V$.}
The redefined physical states ${W'}^{\pm},\,Z'$ and $A'$ are obtained from 
${W'}^a$ and $B'$ in the usual way in terms of the redefined mixing angle.
For instance, $Z'=Z\sqrt{1+C_Z}$. Since the interactions between the gauge 
bosons and fermions are unchanged, the $A'$ field is by definition the photon 
field and the redefined coupling $e' = g'_1g'_2/\sqrt{{g'_1}^2+{g'_2}^2}$ is 
by definition the electromagnetic coupling constant. This is equivalent to
performing finite renormalizations in order to absorb the non-local 
contributions to the photon wave function and the electromagnetic charge. 
So, the combined effect of all non-local interactions amounts to the effective 
replacements given in Eq.~(\ref{examples/YM/upshot}). These effective 
replacements can be extended to the longitudinal sector by simply rescaling the
gauge parameter in the gauge-fixing part of the Lagrangian. For instance,
with the rescaling $\xi_{_W} = \xi'_{_W}/(1+\Sigma_2)$ the dressed $W$-boson 
propagator in Eq.~(\ref{prop/W}) becomes
\be
  P^{WW}_{\mu\nu}(q,\xi_{_W}) 
  = 
  \frac{-i}{[1+\Sigma_2]\,[q^2-{M'_W\!\!}^2]}\,
  \biggl( g_{\mu\nu} - \frac{q_{\mu}q_{\nu}}{q^2} \biggr)
  - \frac{i\xi'_{_W}}{[1+\Sigma_2]\,[q^2-\xi'_{_W}{M'_W\!\!}^2]}\,
  \frac{q_{\mu}q_{\nu}}{q^2},
\ee
where the factor $1/(1+\Sigma_2)$ can be absorbed into the $W$-boson fields
according to Eq.~(\ref{redef}). This simple example for the gauge bosons 
coincides with the complex-mixing-angle procedure that was adopted in 
Ref.~\cite{DDRrad} for calculating the radiative processes 
$e^+e^- \to 4f\gamma$. 

The appeal of the above-discussed special examples lies in the simplicity of 
the net prescriptions that follow from the effective Lagrangians, allowing a
straightforward implementation into the existing Monte Carlo programs. In spite
of the simplicity, nevertheless a reasonably good description of the 
unstable-particle resonances can be achieved.
In case of a more rigorous treatment of unstable particles, involving a proper 
energy dependence of the absorptive parts of the self-energies, one is forced
to take into account the full extent of the effective Lagrangians. Or, in other
words, one has to properly take into account the relevant sets of 
gauge-restoring multi-particle interactions (see e.g.~the Feynman rules listed
in App.~\ref{app:b}).


\section{Conclusions and outlook}

In this paper we have introduced a method that offers the
possibility of performing gauge-invariant tree-level calculations
with unstable particles in intermediate states. To this end non-local
gauge-invariant Lagrangians are introduced, which allow the gauge-invariant
resummation of self-energies and therefore give rise to dressed (regular)
propagators for unstable particles. Certainly for practical applications 
the resummed self-energies will be taken from the underlying gauge theory,
but in principle the choice is arbitrary in our approach. For every choice the 
gauge restoring vertices will be different. This leaves open the possibility 
of studying {\it ad hoc} methods for implementing the decay widths of the 
unstable particles, like the fixed-width scheme. From the non-local Lagrangian 
one obtains in general an infinite set of multi-point vertices. 
These vertices provide an explicit solution of the full set of 
ghost-free Ward identities and thereby restore the gauge invariance of the 
resummed amplitudes. For a given multi-particle process only a limited number 
of those vertices contribute. In the paper we have given the derivation of the 
multi-point vertices from the non-local Lagrangians, and we have explicitly 
listed all relevant modifications of the Standard Model vertices for up to 
four external particles. These modified vertices are related to the unstable 
gauge bosons, the Higgs particle and the top-quark, which all occur in the 
electroweak/QCD calculations for present and future collider experiments. 
It should be kept in mind that there are other multi-point vertices that would 
also lead to gauge-invariant amplitudes. In other words, the vertices are not 
unique, but our prescription gives in a minimal way a set of vertices that 
restores gauge invariance.

Usually one restricts the final-state particles in a process to stable
particles, i.e.~fermions, photons and gluons. The final-state fermions can be 
either massive or massless. In our approach this poses no problem, since the 
calculation remains gauge-invariant in either case. The vertices given in 
App.~\ref{app:b} allow gauge-invariant calculations for unstable-particle 
processes like $e^{+}e^{-}/q\bar{q}\,/gg\to 4f\gamma,\,4fg,\,6f$. 
Many of the present-day unstable-particle production processes lead to these 
final states. Examples are $W^{+}W^{-}\gamma$ production at LEP2 and $t\bar{t}$
production at the Tevatron. For the latter process gluon radiation could
also be of practical importance. In that case one should
extend the list in App.~\ref{app:b} and add the vertices that contribute 
to a final state with six fermions and one gluon.
For instance, a 3-gluon--$t\bar{t}$ vertex would arise.

Although this paper was primarily motivated by the phenomenological
need to perform sensible tree-level calculations with unstable
particles, other applications seem possible.

One possible application could be the gauge-invariant resummation of gluon 
propagators in QCD calculations. In this way part of the higher-order 
corrections can be taken into account in a gauge-invariant way. The effect of 
this resummation on multiparton amplitudes can now be investigated
using our method. In a similar way one could study the resummation 
of the electroweak gauge-boson propagators in terms of running (effective)
couplings.

Another intriguing question is whether the non-local Lagrangian technique
could be used to construct a gauge-invariant bosonic self-energy
by adding gauge-restoring parts from vertex and box diagrams 
to a non-gauge-invariant self-energy. 
In other words, could the method of non-local Lagrangians be used 
to carry out the pinch technique?

Another issue is whether one could use the propagators and
vertices derived in the paper to perform quantum loop corrections. 
To our knowledge this remains an open question.


\appendix
\renewcommand{\theequation}{\thesection .\arabic{equation}}

\section{Expressions for the path-ordered tensor-functions}
\setcounter{equation}{0}
\label{app:a}

In this appendix we derive explicit expressions for the path-ordered 
tensor-functions $\A_n$, introduced in Sect.~\ref{sec:top}. We start off the 
derivation by solving a set of scalar recursion relations. The simplest one is 
defined as 
\be
  \X_n(l',l) = l^2 \X_{n-1}(l',l), 
  \qquad
  \X_0(l',l)= (2\pi)^4\,\delta^{(4)}(l'+l),
\ee
which has the trivial solution 
\be
  \X_{n}(l',l) = l^{2n}\,(2\pi)^4\,\delta^{(4)}(l'+l).
\ee
Note that such a scalar function translates directly into a non-local 
coefficient when the summation over $n$ is performed:
$\sum_{n=0}^{\infty}\frac{1}{n!}\,\tilde{\Sigma}_{\sss{NL}}^{(n)}(0)\,l^{2n} 
= \tilde{\Sigma}_{\sss{NL}}(l^2)$. For the path-ordered tensor-functions we 
will need to solve the following more general set of scalar recursion 
relations:
\begin{eqnarray}
  \X_n(l',l|q_1,..\,,q_k) &=& l^2 \X_{n-1}(l',l|q_1,..\,,q_k)
                          + \X_{n-1}(l',l+q_k|q_1,..\,,q_{k-1}) 
                          \hspace*{4ex}(k\ge 2), \nonumber \\[2mm]
  \X_n(l',l|q_1)      &=& l^2 \X_{n-1}(l',l|q_1) + \X_{n-1}(l',l+q_1),
\end{eqnarray}
with the base of the recursion given by
\be
  \X_0(l',l|q_1,..\,,q_k) = \X_0(l',l|q_1) = 0.
\ee
The solutions of these recursion relations read
\be
\label{app:X/solution}
  \X_n(l',l|q_1,..\,,q_k)
  =
  (2\pi)^4\,\delta^{(4)}(l'+P_1)\,\sum\limits_{i=1}^{k+1}
  \frac{P_i^{2n}}{\prod\limits_{j=1,\,j\neq i}^{k+1}
                  \Bigl( P_i^2-P_j^2 \Bigr)},
\ee
with 
\be
  P_i = l+\sum\limits_{j=i}^k q_j \hspace*{5ex} (i\le k)
  \qquad \mbox{and} \qquad
  P_{k+1} = l.
\ee
Since $P_i^{2n}-l^2 P_i^{2n-2} = P_i^{2n-2}\,(P_i^2-P_{k+1}^2)$,
one can easily verify that Eq.~(\ref{app:X/solution}) indeed represents  a set
of solutions. Note again that each term occurring in these solutions 
translates directly into a non-local coefficient when the summation over $n$ 
is performed [$P_i^{2n} \to \tilde{\Sigma}_{\sss{NL}}(P_i^2)$].

After these preparations we can now turn to the path-ordered tensor-functions
\be
  \tilde{\A}_n^{\mu_1..\mu_k}(l',l|q_1,..\,,q_k)
  =
  (-i)^k\!\int d^4\xi\,d^4\tau\,\delta^{(4)}(\xi-\tau)\,e^{-il'\cdot\xi}
  (-\partial^2_{\tau})^n\,e^{-il\cdot\tau}\,\prod\limits_{j=1}^k
  \,\,\int\limits_{\omega_{j\!-\!1}}^{\tau}d\omega_j^{\mu_j}\,
  e^{-iq_j\cdot\omega_j},
\ee
with $\omega_0 = \xi$. By working out one of the $(-\partial_{\tau}^2)$ 
operators one arrives at the following set of tensor recursion relations:
\begin{eqnarray}
  \tilde{\A}_n(l',l)                         &=& l^2 \tilde{\A}_{n-1}(l',l),
       \\[2mm]
  \tilde{\A}_n^{\mu_1}(l',l|q_1)             &=&
         l^2 \tilde{\A}_{n-1}^{\mu_1}(l',l|q_1)
       + (2l+q_1)^{\mu_1}\,\tilde{\A}_{n-1}(l',l+q_1), 
       \nonumber \\[2mm]
  \tilde{\A}_n^{\mu_1\mu_2}(l',l|q_1,q_2)    &=&
         l^2 \tilde{\A}_{n-1}^{\mu_1\mu_2}(l',l|q_1,q_2)
       + (2l+q_2)^{\mu_2}\,\tilde{\A}_{n-1}^{\mu_1}(l',l+q_2|q_1)
       \nonumber \\[2mm]
                                             & &
       +\,g^{\mu_1\mu_2}\tilde{\A}_{n-1}(l',l+q_1+q_2), 
       \nonumber \\[2mm]
  \tilde{\A}_n^{\mu_1..\mu_k}(l',l|q_1,..\,,q_k) &=&
         l^2 \tilde{\A}_{n-1}^{\mu_1..\mu_k}(l',l|q_1,..\,,q_k)
       + (2l+q_k)^{\mu_k}\,\tilde{\A}_{n-1}^{\mu_1..\mu_{k-1}}
                                     (l',l+q_k|q_1,..\,,q_{k-1})
       \nonumber \\[2mm]
                                             & &
       +\,g^{\mu_{k-1}\mu_k}\tilde{\A}_{n-1}^{\mu_1..\mu_{k-2}}
                                      (l',l+q_{k-1}+q_k|q_1,..\,,q_{k-2}) 
       \hspace*{14ex} (k\ge 3). \nonumber
\end{eqnarray}                      
The base of the recursion is given by the relations
\be
  \tilde{\A}_0^{\mu_1..\mu_k}(l',l|q_1,..\,,q_k) = 0 \hspace*{5ex} (k\ge 1)
  \qquad \mbox{and} \qquad
  \tilde{\A}_0(l',l) = (2\pi)^4\,\delta^{(4)}(l'+l).
\ee
Evidently $\tilde{\A}_n(l',l)$ is identical to $\X_n(l',l)$. The other
tensor-functions can also be expressed in a straightforward way in terms of 
the afore-mentioned solutions of the scalar recursion relations:
\begin{eqnarray}
  \lefteqn{\hspace*{-3ex}\tilde{\A}_n^{\mu_1..\mu_k}(l',l|q_1,..\,,q_k) =
      \X_n(l',l|q_1,..\,,q_k)\,Q_1^{\mu_1}\cdots \,Q_k^{\mu_k}}
      \nonumber \\[2mm]
                                             & &
    +\,\sum\limits_{m=1}^{k-1} \X_n(l',l|q_1,..\,,q_{m-1},q_m+q_{m+1},
                                         q_{m+2},..\,,q_k)
      \,Q_1^{\mu_1}\cdots \,Q_{m-1}^{\mu_{m-1}}\,g^{\mu_m\mu_{m+1}}\,
      Q_{m+2}^{\mu_{m+2}}\cdots \,Q_k^{\mu_k}
      \nonumber \\[2mm]
                                             & &
    +\,\mbox{\ two insertions of the metric tensor $g$\ } + \ \cdots\,,
\end{eqnarray}   
with
\be
  Q_i^{\mu_i} = P_i^{\mu_i} + P_{i+1}^{\mu_i}.
\ee
As a final step we insert the explicit solutions (\ref{app:X/solution}):
\begin{eqnarray}
  \tilde{\A}_n^{\mu_1..\mu_k}(l',l|q_1,..\,,q_k) &=&
      (2\pi)^4\,\delta^{(4)}(l'+P_1)\,\sum\limits_{i=1}^{k+1}
      \frac{P_i^{2n}}{\prod\limits_{j=1,\,j\neq i}^{k+1}
                      \Bigl(P_i^2-P_j^2\Bigr)}
      \,\,O_i^{\mu_1..\mu_k}(l|q_1,..\,,q_k),
      \nonumber \\[1mm]
  O_i^{\mu_1..\mu_k}(l|q_1,..\,,q_k)             &=&
      Q_1^{\mu_1}\cdots \,Q_k^{\mu_k}
    + \sum\limits_{m=1}^{k-1} \Bigl( P_i^2-P_{m+1}^2 \Bigr)\,
      Q_1^{\mu_1}\cdots \,Q_{m-1}^{\mu_{m-1}}\,g^{\mu_m\mu_{m+1}}\,
      Q_{m+2}^{\mu_{m+2}}\cdots \,Q_k^{\mu_k}
      \nonumber \\[2mm]
                                             & &
    +\,\mbox{\ two insertions of the metric tensor $g$\ } + \ \cdots
\end{eqnarray}
As a result of the relation $q_{i,\,\mu_i}\,Q_i^{\mu_i} = (P_i^2-P_{i+1}^2)$,
the tensor-functions obey the simple `Ward identities'
\begin{eqnarray}
  q_{1,\,\mu_1}\,\tilde{\A}_n^{\mu_1}(l',l|q_1)                 \!\!&=&\!\!
      \tilde{\A}_n(l',l\!+\!q_1) - \tilde{\A}_n(l'\!+\!q_1,l),
      \\[2mm]
  q_{1,\,\mu_1}\,\tilde{\A}_n^{\mu_1..\mu_k}(l',l|q_1,..\,,q_k) \!\!&=&\!\!
      \tilde{\A}_n^{\mu_2..\mu_k}(l',l|q_1\!+\!q_2,q_3,..\,,q_k)
      - \tilde{\A}_n^{\mu_2..\mu_k}(l'\!+\!q_1,l|q_2,..\,,q_k),
      \nonumber \\[2mm]
  q_{k,\,\mu_k}\,\tilde{\A}_n^{\mu_1..\mu_k}(l',l|q_1,..\,,q_k) \!\!&=&\!\!
      \tilde{\A}_n^{\mu_1..\mu_{k-1}}(l',l\!+\!q_k|q_1,..\,,q_{k-1})
      - \tilde{\A}_n^{\mu_1..\mu_{k-1}}(l',l|q_1,..\,,q_{k-2},q_{k-1}\!+\!q_k),
      \nonumber \\[2mm]
  q_{r,\,\mu_r}\,\tilde{\A}_n^{\mu_1..\mu_k}(l',l|q_1,..\,,q_k) \!\!&=&\!\!
      \tilde{\A}_n^{\mu_1..\mu_{r-1}\mu_{r+1}..\mu_k}
                (l',l|q_1,..\,,q_{r-1},q_r\!+\!q_{r+1},q_{r+2},..\,,q_k)
      \nonumber \\[2mm]
                                                                \!\!&-&\!\! 
      \tilde{\A}_n^{\mu_1..\mu_{r-1}\mu_{r+1}..\mu_k}
                (l',l|q_1,..\,,q_{r-2},q_{r-1}\!+\!q_r,q_{r+1},..\,,q_k)
      \hspace*{3ex} (1<r<k). \nonumber
\end{eqnarray}

As an example we present the explicit solutions for $k=0,1,2$, which are
relevant for the derivation of the two-, three- and four-point interactions 
that are presented in this paper:
\begin{eqnarray}
\label{An/explicit}
  \tilde{\A}_n(l',l)                      &=& 
             (2\pi)^4\,\delta^{(4)}(l'+l)\,l^{2n}, \\[1mm]
  \tilde{\A}_n^{\mu_1}(l',l|q_1)          &=&
             (2\pi)^4\,\delta^{(4)}(l'+l+q_1)\,
             \frac{(l+q_1)^{2n}-l^{2n}}{(l+q_1)^2-l^2}\,(2l+q_1)^{\mu_1},
              \nonumber \\[1mm]
  \tilde{\A}_n^{\mu_1\mu_2}(l',l|q_1,q_2) &=& 
             (2\pi)^4\,\delta^{(4)}(l'+l+q_1+q_2)\,
             \Biggl[ g^{\mu_1\mu_2}\,\frac{(l+q_1+q_2)^{2n}-l^{2n}}
                                          {(l+q_1+q_2)^2-l^2} 
             \nonumber \\[1mm]
                                          & & \hspace*{-10ex}
             +\,(2l+2q_2+q_1)^{\mu_1}\,(2l+q_2)^{\mu_2}
             \biggl\{ \frac{(l+q_1+q_2)^{2n}}{[(l+q_1+q_2)^2-(l+q_2)^2]
                                             [(l+q_1+q_2)^2-l^2]}
             \nonumber \\[1mm]
                                          & & \hspace*{-10ex}
             -\,\frac{(l+q_2)^{2n}}{[(l+q_1+q_2)^2-(l+q_2)^2][(l+q_2)^2-l^2]}
             + \frac{l^{2n}}{[(l+q_1+q_2)^2-l^2][(l+q_2)^2-l^2]} 
             \biggr\} \Biggr]. \nonumber
\end{eqnarray}

\section{Some non-local Feynman rules}
\setcounter{equation}{0}
\label{app:b}

In this appendix we list the non-local contributions to the various three-
and four-point interactions. Whenever possible we will suppress the factor 
$(2\pi)^4$ and the delta-function for momentum conservation. 

First we give the non-local contributions to the 
pure gauge-boson interactions as originating from the non-local actions
$\SNL^{\sss{YM}}$ in Eq.~(\ref{sm/action}) and $\SNL^{\Phi}$ in 
Eq.~(\ref{sm/action/higgs}). We start with the three-point gauge-boson 
interaction:
\begin{eqnarray}
  \begin{picture}(75,45)(0,-2)
    \Photon(0,0)(40,0){2}{5}
    \ArrowLine(25,-12)(26,-12)
    \Line(15,-12)(25,-12)
    \Photon(40,0)(60,30){2}{5}
    \ArrowLine(57,8)(55,5)
    \Line(57,8)(62,15.5)
    \Photon(60,-30)(40,0){2}{5}
    \ArrowLine(57,-8)(55,-5)
    \Line(57,-8)(62,-15.5)
    \GCirc(40,0){3}{0}
    \Text(0,8)[lb]{$V_1,\,\mu_1$}
    \Text(0,-8)[lt]{$q_1$}
    \Text(60,35)[cb]{$V_2,\,\mu_2$}
    \Text(65,20)[lb]{$q_2$}
    \Text(60,-35)[ct]{$V_3,\,\mu_3$}
    \Text(65,-20)[lt]{$q_3$}
  \end{picture} 
  \hspace*{2ex} 
  &:& ig_2\,\Biggl\{ A_2\,\,\sum_{n=0}^{\infty} \frac{1}{n!}\,
                     \tilde{\Sigma}_2^{(n)}(0)\,\sum\limits_{\sss{perm}}
                     \epsilon^{jkl}\,
                     \tilde{V}_{\sss{NL},\,n}^{\mu_j\mu_k\mu_l}(q_j,q_k,q_l)
      \hspace*{7ex} \nonumber \\[3mm]
  & & \hphantom{ig_2 a} +\,A^{\mu_1,\,\mu_2\mu_3}(q_1)\,\Bigl[ 
                              A_{31}\,\tilde{\Sigma}_3(q_1^2)
                            + A_{41}\,\tilde{\Sigma}_4(q_1^2) \Bigr]
            \Biggr\},
\end{eqnarray}
with the various couplings given by 
\be
  \begin{array}{|l|c|c|c|}
    \hline
    V_1V_2V_3         & A_2  & A_{31}    & A_{41} \\
    \hline & & & \\[-4.5mm] 
    ZW^{+}W^{-}       & -c_w & s_w^2/c_w & -c_w   \\
    \gamma W^{+}W^{-} & s_w  & s_w       & s_w    \\
    \hline
  \end{array}
\ee
The sum over the permutations involves all permutations $(j,k,l)$ of the 
labels $(1,2,3)$. Now we can make use of Eqs.~(\ref{ww/k-point}) and 
(\ref{An/explicit}) to arrive at
\begin{eqnarray}
  \lefteqn{\hspace*{-2.0cm} \sum_{n=0}^{\infty} \frac{1}{n!}\,
            \tilde{\Sigma}_2^{(n)}(0)\,
            \tilde{V}_{\sss{NL},\,n}^{\mu_1\mu_2\mu_3}(q_1,q_2,q_3)
            = \sum_{n=0}^{\infty} \frac{1}{n!}\,\tilde{\Sigma}_2^{(n)}(0)\,
            \biggl\{ \frac{1}{2}\,T^{\mu_1\mu_2}(q_1,q_2)\,
                     \tilde{\A}_n^{\mu_3}(q_1,q_2|q_3)}
  \nonumber \\[1mm]
  & & +\,\frac{1}{4}\,A^{\mu_1,\,\mu_2\mu_3}(q_1)\,\tilde{\A}_n(q_1,q_2+q_3)
      - \frac{1}{4}\,A^{\mu_2,\,\mu_1\mu_3}(q_2)\,\tilde{\A}_n(q_1+q_3,q_2)
            \biggr\}
      \nonumber \\[2mm]
  &\to & \tilde{\Sigma}_2(q_1^2)\, \biggl\{ 
         \frac{1}{2}\,A^{\mu_1,\,\mu_2\mu_3}(q_1)
         + \frac{(2q_1+q_3)^{\mu_3}}{(q_1+q_3)^2-q_1^2}\,
           T^{\mu_1\mu_2}(q_1,q_2) \biggr\}.
\end{eqnarray}
In the last step we have compactified the expression by exploiting the fact 
that the summation over all permutations has to be performed and that 
$\epsilon^{jkl}$ is totally antisymmetric. Moreover, the factor 
$(2\pi)^4\,\delta^{(4)}(q_1+q_2+q_3)$ has been suppressed.

The four-point gauge-boson interaction is modified according to
\begin{eqnarray}
\hspace*{3ex}
\begin{picture}(75,50)(0,-2)
{
   \Photon(0,30)(30,0){2}{5}
   \ArrowLine(7,8)(10,5)
   \Line(7,8)(0,15)
   \Photon(0,-30)(30,0){2}{5}
   \ArrowLine(7,-8)(10,-5)
   \Line(7,-8)(0,-15)
   \Photon(30,0)(60,30){2}{5}
   \ArrowLine(53,8)(50,5)
   \Line(53,8)(60,15)
   \Photon(60,-30)(30,0){2}{5}
   \ArrowLine(53,-8)(50,-5)
   \Line(53,-8)(60,-15)
   \GCirc(30,0){3}{0}
   \Text(0,35)[cb]{$V_1,\,\mu_1$}
   \Text(-5,19)[rb]{$q_1$}
   \Text(0,-35)[ct]{$V_2,\,\mu_2$} 
   \Text(-5,-19)[rt]{$q_2$}
   \Text(60,35)[cb]{$V_3,\,\mu_3$}
   \Text(65,19)[lb]{$q_3$}
   \Text(60,-35)[ct]{$V_4,\,\mu_4$}
   \Text(65,-19)[lt]{$q_4$} 
}
\end{picture}
\hspace*{1.5ex}
  &:& ig_2^2\,\Biggl\{ B_2\,\,\sum_{n=0}^{\infty} \frac{1}{n!}\,
          \tilde{\Sigma}_2^{(n)}(0)\,\sum\limits_{\sss{perm}}\eta_{jklm}\,
          \tilde{V}_{\sss{NL},\,n}^{\mu_j\mu_k\mu_l\mu_m}(q_j,q_k,q_l,q_m) 
      \hspace*{7ex} \nonumber \\[3mm]
  & & \hphantom{ig_2 a} +\,B_{413}\,\tilde{\Sigma}_4([q_1+q_3]^2)\,\Bigl( 
            g^{\mu_1\mu_2}g^{\mu_3\mu_4} - g^{\mu_1\mu_4}g^{\mu_2\mu_3} \Bigr)
      \nonumber \\[3mm]
  & & \hphantom{ig_2 a} +\,B_{414}\,\tilde{\Sigma}_4([q_1+q_4]^2)\,\Bigl( 
            g^{\mu_1\mu_2}g^{\mu_3\mu_4} - g^{\mu_1\mu_3}g^{\mu_2\mu_4} \Bigr)
              \Biggr\},
  \vspace*{7ex}
\end{eqnarray}
with the various couplings given by 
\vspace*{1ex}
\be
  \begin{array}{|l|c|c|c|c|c|c|}
    \hline
    V_1V_2V_3V_4           & B_2     & B_{413} & B_{414} \\
    \hline & & & \\[-4.5mm]
    W^{+}W^{-}ZZ           & -c_w^2  & 0       & 0       \\
    W^{+}W^{-}Z\gamma      & s_w c_w & 0       & 0       \\
    W^{+}W^{-}\gamma\gamma & -s_w^2  & 0       & 0       \\
    W^{+}W^{+}W^{-}W^{-}   & 1       & 1       & 1       \\ 
    \hline
  \end{array}
  \vspace*{1ex}
\ee
The sum over the permutations involves all permutations $(j,k,l,m)$ of the 
labels $(1,2,3,4)$ and
\be
  \eta_{jklm} = \left\{ \begin{array}{cl}
                          0 & \mbox{ if \ }(j,k,l,m) \ = (1,3,2,4) \,,\,
                              (4,2,3,1) \mbox{ \ or any \ }
                              1\leftrightarrow 2 \,,\, 3\leftrightarrow 4 
                              \mbox{ \ permutation} \\
                         +1 & \mbox{ if \ }(j,k,l,m) \ = (1,3,4,2) \,,\,
                              (4,2,1,3) \mbox{ \ or any \ }  
                              1\leftrightarrow 2 \,,\, 3\leftrightarrow 4 
                              \mbox{ \ permutation} \ \ .\\
                         -1 & \mbox{ if \ }(j,k,l,m) \ = (1,2,3,4) \,,\,
                              (3,4,1,2) \mbox{ \ or any \ } 
                              1\leftrightarrow 2 \,,\, 3\leftrightarrow 4 
                              \mbox{ \ permutation}
                        \end{array} \right. 
\ee 
Now we can make use of Eqs.~(\ref{ww/k-point}) and (\ref{An/explicit}) to 
arrive at
\begin{eqnarray}
  \lefteqn{\hspace*{-0.5cm} \sum_{n=0}^{\infty} \frac{1}{n!}\,
            \tilde{\Sigma}_2^{(n)}(0)\,
            \tilde{V}_{\sss{NL},\,n}^{\mu_1\mu_2\mu_3\mu_4}(q_1,q_2,q_3,q_4)
            = \sum_{n=0}^{\infty} \frac{1}{n!}\,\tilde{\Sigma}_2^{(n)}(0)\,
            \biggl\{ \frac{1}{2}\,T^{\mu_1\mu_2}(q_1,q_2)\,
                     \tilde{\A}_n^{\mu_3\mu_4}(q_1,q_2|q_3,q_4)}
  \nonumber \\[1mm]
  & & +\,\frac{1}{4}\,A^{\mu_1,\,\mu_2\mu_4}(q_1)\,
         \tilde{\A}_n^{\mu_3}(q_1,q_2+q_4|q_3)
      -  \frac{1}{4}\,A^{\mu_2,\,\mu_1\mu_3}(q_2)\,
         \tilde{\A}_n^{\mu_4}(q_1+q_3,q_2|q_4)
      \nonumber \\[2mm]
  & & -\,\frac{1}{4}\,g^{\mu_1\mu_2}\,g^{\mu_3\mu_4}\,
         \tilde{\A}_n(q_1+q_3,q_2+q_4)
            \biggr\}
      \nonumber \\[2mm]
  &\to & \tilde{\Sigma}_2(q_1^2)\,\frac{T^{\mu_1\mu_2}(q_1,q_2)}{q_1^2-q_2^2}\,
         \biggl\{ g^{\mu_3\mu_4} + \frac{(2q_2+q_4)^{\mu_4}(2q_1+q_3)^{\mu_3}}
                                        {(q_1+q_3)^2-q_1^2} 
         \biggr\}
         \nonumber \\[2mm]
  & & -\,\frac{1}{4}\,\tilde{\Sigma}_2([q_1+q_3]^2)\,\biggl\{
      g^{\mu_1\mu_2}g^{\mu_3\mu_4} 
      + 2\,\frac{(2q_1+q_3)^{\mu_3}}{(q_1+q_3)^2-q_1^2}\,
        \frac{(2q_2+q_4)^{\mu_4}}{(q_2+q_4)^2-q_2^2}\,T^{\mu_1\mu_2}(q_1,q_2)
                                                              \biggr\}
         \nonumber \\[2mm]
  & & +\,\frac{1}{2}\,\frac{(2q_1+q_3)^{\mu_3}}{(q_1+q_3)^2-q_1^2}\,
      A^{\mu_1,\,\mu_2\mu_4}(q_1)\,
      \biggl\{ \tilde{\Sigma}_2(q_1^2) - \tilde{\Sigma}_2([q_1+q_3]^2) 
      \biggr\}. 
\end{eqnarray}
In the last step we have again exploited the symmetry properties of the 
summation over all permutations.

The non-local action $\SNL^{\Phi}$ in Eq.~(\ref{sm/action/higgs}) also contains
explicit interactions between gauge bosons and physical/unphysical Higgs 
bosons. The contribution to the interaction between one scalar particle and 
two gauge bosons reads
\be
  \begin{picture}(75,45)(0,-2)
    \DashLine(0,0)(40,0){4}
    \ArrowLine(25,-12)(26,-12)
    \Line(15,-12)(25,-12)
    \Photon(40,0)(60,30){2}{5}
    \ArrowLine(57,8)(55,5)
    \Line(57,8)(62,15.5)
    \Photon(60,-30)(40,0){2}{5}
    \ArrowLine(57,-8)(55,-5)
    \Line(57,-8)(62,-15.5)
    \GCirc(40,0){3}{0}
    \Text(0,8)[lb]{$S$}
    \Text(0,-8)[lt]{$q$}
    \Text(60,35)[cb]{$V_1,\,\mu_1$}
    \Text(65,20)[lb]{$q_1$}
    \Text(60,-35)[ct]{$V_2,\,\mu_2$}
    \Text(65,-20)[lt]{$q_2$}
  \end{picture} 
  \hspace*{2ex} 
  :\ \frac{ig_2}{M_W}\,T^{\mu_1\mu_2}(q_1,q_2)\,
     \biggl\{ \frac{s_w}{c_w}\,\Bigl[ C_{31}\,\tilde{\Sigma}_3(q_1^2) 
                                    + C_{32}\,\tilde{\Sigma}_3(q_2^2)
                               \Bigr]
            + C_{41}\,\tilde{\Sigma}_4(q_1^2) 
            + C_{42}\,\tilde{\Sigma}_4(q_2^2)
     \biggr\},
  \vspace*{7ex}
\ee
with the various couplings given by 
\vspace*{1ex}
\be
  \begin{array}{|l|c|c|c|c|}
    \hline
    SV_1V_2                  & C_{31}   & C_{32}   & C_{41}   & C_{42}   \\
    \hline & & & & \\[-4.5mm]
    HZZ                      & -s_w c_w & -s_w c_w & c_w^2    & c_w^2    \\
    HZ\gamma                 & s_w^2    & -c_w^2   & -s_w c_w & -s_w c_w \\
    H\gamma\gamma            & s_w c_w  & s_w c_w  & s_w^2    & s_w^2    \\ 
    \phi^{\mp}ZW^{\pm}       & s_w      & 0        & -c_w     & 0        \\
    \phi^{\mp}\gamma W^{\pm} & c_w      & 0        & s_w      & 0        \\
    \hline
  \end{array}
  \vspace*{1ex}
\ee 
For the interaction between two scalar particles and two gauge bosons we obtain
\begin{eqnarray}
\hspace*{2ex}
\begin{picture}(75,50)(0,-2)
{
   \Photon(0,30)(30,0){2}{5}
   \ArrowLine(7,8)(10,5)
   \Line(7,8)(0,15)
   \Photon(0,-30)(30,0){2}{5}
   \ArrowLine(7,-8)(10,-5)
   \Line(7,-8)(0,-15)
   \DashLine(30,0)(60,30){4}
   \ArrowLine(53,8)(50,5)
   \Line(53,8)(60,15)
   \DashLine(60,-30)(30,0){4}
   \ArrowLine(53,-8)(50,-5)
   \Line(53,-8)(60,-15)
   \GCirc(30,0){3}{0}
   \Text(0,35)[cb]{$V_1,\,\mu_1$}
   \Text(-5,19)[rb]{$q_1$}
   \Text(0,-35)[ct]{$V_2,\,\mu_2$} 
   \Text(-5,-19)[rt]{$q_2$}
   \Text(60,35)[cb]{$S_1$}
   \Text(65,19)[lb]{$q_3$}
   \Text(60,-35)[ct]{$S_2$}
   \Text(65,-19)[lt]{$q_4$} 
}
\end{picture}
\hspace*{2ex}
  &:& \frac{ig_2^2}{2M_W^2}\,T^{\mu_1\mu_2}(q_1,q_2)\,
      \biggl\{ \frac{s_w}{c_w}\,\Bigl[ D_{31}\,\tilde{\Sigma}_3(q_1^2) 
                                     + D_{32}\,\tilde{\Sigma}_3(q_2^2)
                                \Bigr]
             + D_{41}\,\tilde{\Sigma}_4(q_1^2) 
      \hspace*{7ex} \nonumber \\[3mm]
  & & \hspace*{10ex} 
             +\,D_{42}\,\tilde{\Sigma}_4(q_2^2)
             + 2\,D_{413}\,\tilde{\Sigma}_4([q_1+q_3]^2) 
             + 2\,D_{414}\,\tilde{\Sigma}_4([q_1+q_4]^2)            
     \biggr\}, \nonumber \\ 
\end{eqnarray}
\vspace*{-1ex}
with the various couplings given by 
\vspace*{1ex} 
\be
  \begin{array}{|l|c|c|c|c|c|c|}
    \hline
    V_1V_2S_1S_2                       & D_{31}   & D_{32}   & D_{41} 
                                       & D_{42}   & D_{413}  & D_{414}  \\
    \hline & & & & & & \\[-4.5mm] 
    ZZHH                               & -s_w c_w & -s_w c_w & c_w^2 
                                       & c_w^2    & c_w^2    & c_w^2    \\
    Z\gamma HH                         & s_w^2    & -c_w^2   & -s_w c_w 
                                       & -s_w c_w & -s_w c_w & -s_w c_w \\
    \gamma\gamma HH                    & s_w c_w  & s_w c_w  & s_w^2 
                                       & s_w^2    & s_w^2    & s_w^2    \\ 
    ZZ\chi\chi                         & -s_w c_w & -s_w c_w & c_w^2 
                                       & c_w^2    & 0        & 0        \\
    Z\gamma\chi\chi                    & s_w^2    & -c_w^2   & -s_w c_w 
                                       & -s_w c_w & 0        & 0        \\
    \gamma\gamma\chi\chi               & s_w c_w  & s_w c_w  & s_w^2 
                                       & s_w^2    & 0        & 0        \\ 
    ZZ\phi^+\phi^-                     & s_w c_w  &  s_w c_w & -c_w^2 
                                       & -c_w^2   & 0        & 0        \\
    Z\gamma\phi^+\phi^-                & -s_w^2   & c_w^2    & s_w c_w 
                                       & s_w c_w  & 0        & 0        \\  
    \gamma\gamma\phi^+\phi^-           & -s_w c_w & -s_w c_w & -s_w^2 
                                       & -s_w^2   & 0        & 0        \\
    ZW^{\pm}\phi^{\mp}H                & s_w      & 0        & -c_w
                                       & 0        & 0        & -c_w     \\
    \gamma W^{\pm}\phi^{\mp}H          & c_w      & 0        & s_w
                                       & 0        & 0        & s_w      \\
    ZW^{\pm}\phi^{\mp}\chi             & \pm is_w & 0        & \mp ic_w
                                       & 0        & 0        & 0        \\     
    \gamma W^{\pm}\phi^{\mp}\chi       & \pm ic_w & 0        & \pm is_w
                                       & 0        & 0        & 0        \\
    W^{+}W^{-}\phi^+\phi^-             & 0        & 0        & 0
                                       & 0        & 0        & 1        \\
    W^{\pm}W^{\pm}\phi^{\mp}\phi^{\mp} & 0        & 0        & 0
                                       & 0        & 1        & 1        \\
    \hline
  \end{array}
  \vspace*{1ex}
\ee
In addition new interactions emerge between one scalar particle and three 
gauge bosons:
\begin{eqnarray}
\hspace*{3ex}
\begin{picture}(75,50)(0,-2)
{
   \Photon(0,30)(30,0){2}{5}
   \ArrowLine(7,8)(10,5)
   \Line(7,8)(0,15)
   \Photon(0,-30)(30,0){2}{5}
   \ArrowLine(7,-8)(10,-5)
   \Line(7,-8)(0,-15)
   \Photon(30,0)(60,30){2}{5}
   \ArrowLine(53,8)(50,5)
   \Line(53,8)(60,15)
   \DashLine(60,-30)(30,0){4}
   \ArrowLine(53,-8)(50,-5)
   \Line(53,-8)(60,-15)
   \GCirc(30,0){3}{0}
   \Text(0,35)[cb]{$V_1,\,\mu_1$}
   \Text(-5,19)[rb]{$q_1$}
   \Text(0,-35)[ct]{$V_2,\,\mu_2$} 
   \Text(-5,-19)[rt]{$q_2$}
   \Text(60,35)[cb]{$V_3,\,\mu_3$}
   \Text(65,19)[lb]{$q_3$}
   \Text(60,-35)[ct]{$S$}
   \Text(65,-19)[lt]{$q_4$} 
}
\end{picture}
\hspace*{1.5ex}
  &:& \frac{ig_2^2}{M_W}\,
      \Biggl\{ A^{\mu_1,\,\mu_2\mu_3}(q_1)\,\biggl[ 
                  \frac{s_w}{c_w}\,E_{31}\,\tilde{\Sigma}_3(q_1^2) 
                  + E_{41}\,\tilde{\Sigma}_4(q_1^2)
                  + E_{414}\,\tilde{\Sigma}_4([q_1+q_4]^2)
                                            \biggr]
      \hspace*{1ex} \nonumber \\[3mm]
  & & \hphantom{\frac{ig_2^2}{M_W}\hspace*{-1.3ex}}
            +\,A^{\mu_2,\,\mu_1\mu_3}(q_2)\,\biggl[ 
                  \frac{s_w}{c_w}\,E_{32}\,\tilde{\Sigma}_3(q_2^2) 
                  + E_{42}\,\tilde{\Sigma}_4(q_2^2)
                  + E_{413}\,\tilde{\Sigma}_4([q_1+q_3]^2)
                                            \biggr] 
      \Biggr\}, \nonumber \\
\end{eqnarray}
with the various couplings given by 
\vspace*{1ex}
\be
  \begin{array}{|l|c|c|c|c|c|c|}
    \hline
    V_1V_2V_3S                      & E_{31}      & E_{32}      & E_{41} 
                                    & E_{42}      & E_{413}     & E_{414}    \\
    \hline & & & & & & \\[-4.5mm] 
    ZZW^{\pm}\phi^{\mp}             & \mp s_w c_w & \mp s_w c_w & \pm c_w^2 
                                    & \pm c_w^2   & 0           & 0          \\
    Z\gamma W^{\pm}\phi^{\mp}       & \pm s_w^2   & \mp c_w^2   & \mp s_w c_w
                                    & \mp s_w c_w & 0        & 0             \\
    \gamma\gamma W^{\pm}\phi^{\mp}  & \pm s_w c_w & \pm s_w c_w & \pm s_w^2 
                                    & \pm s_w^2   & 0           & 0          \\
    ZW^{+}W^{-}H                    & s_w         & 0           & -c_w
                                    & 0           & 0           & -c_w       \\
    \gamma W^{+}W^{-}H              & c_w         & 0           & s_w
                                    & 0           & 0           & s_w        \\
    W^{\pm}W^{\pm}W^{\mp}\phi^{\mp} & 0           & 0           & 0
                                    & 0           & \pm 1       & \pm 1      \\
    \hline
  \end{array}
  \vspace*{1ex}
\ee

The non-local action $\SNL^H$ in Eq.~(\ref{sm/NL/higgs}) modifies the three- 
and four-point interactions between the physical and unphysical Higgs bosons. 
The contribution to the scalar three-point interaction is
\be
  \begin{picture}(75,45)(0,-2)
    \DashLine(0,0)(40,0){4}
    \ArrowLine(25,-12)(26,-12)
    \Line(15,-12)(25,-12)
    \DashLine(40,0)(60,30){4}
    \ArrowLine(57,8)(55,5)
    \Line(57,8)(62,15.5)
    \DashLine(60,-30)(40,0){4}
    \ArrowLine(57,-8)(55,-5)
    \Line(57,-8)(62,-15.5)
    \GCirc(40,0){3}{0}
    \Text(0,8)[lb]{$S_1$}
    \Text(0,-8)[lt]{$q_1$}
    \Text(60,35)[cb]{$S_2$}
    \Text(65,20)[lb]{$q_2$}
    \Text(60,-35)[ct]{$S_3$}
    \Text(65,-20)[lt]{$q_3$}
  \end{picture} 
  \hspace*{2ex} 
  :\ \frac{i}{v}\,\Bigl[ C_1^{(3S)}\,\tilde{\Sigma}_H(q_1^2) 
                       + C_2^{(3S)}\,\tilde{\Sigma}_H(q_2^2)
                       + C_3^{(3S)}\,\tilde{\Sigma}_H(q_3^2) \Bigr],
  \vspace*{7ex}
\ee
with the various couplings given by 
\vspace*{1ex} 
\be
  \begin{array}{|l|c|c|c|}
    \hline & & & \\[-4.5mm]
    S_1S_2S_3     & C_1^{(3S)} & C_2^{(3S)} & C_3^{(3S)} \\
    \hline 
    HHH           & 1          & 1          & 1          \\
    H\chi\chi     & 1          & 0          & 0          \\
    H\phi^+\phi^- & 1          & 0          & 0          \\
    \hline
  \end{array}
  \vspace*{1ex}
\ee  
The contribution to the scalar four-point interaction is
\be
\hspace*{2ex}
\begin{picture}(75,45)(0,-2)
{
   \DashLine(0,30)(30,0){4}
   \ArrowLine(7,8)(10,5)
   \Line(7,8)(0,15)
   \DashLine(0,-30)(30,0){4}
   \ArrowLine(7,-8)(10,-5)
   \Line(7,-8)(0,-15)
   \DashLine(30,0)(60,30){4}
   \ArrowLine(53,8)(50,5)
   \Line(53,8)(60,15)
   \DashLine(60,-30)(30,0){4}
   \ArrowLine(53,-8)(50,-5)
   \Line(53,-8)(60,-15)
   \GCirc(30,0){3}{0}
   \Text(0,35)[cb]{$S_1$}
   \Text(-5,19)[rb]{$q_1$}
   \Text(0,-35)[ct]{$S_2$} 
   \Text(-5,-19)[rt]{$q_2$}
   \Text(60,35)[cb]{$S_3$}
   \Text(65,19)[lb]{$q_3$}
   \Text(60,-35)[ct]{$S_4$}
   \Text(65,-19)[lt]{$q_4$} 
}
\end{picture}
\hspace*{2ex}
  :\ \frac{i}{v^2}\,\Bigl[ C_{12}^{(4S)}\,\tilde{\Sigma}_H([q_1+q_2]^2) 
                         + C_{13}^{(4S)}\,\tilde{\Sigma}_H([q_1+q_3]^2)
                         + C_{14}^{(4S)}\,\tilde{\Sigma}_H([q_1+q_4]^2) \Bigr],
  \vspace*{6ex}
\ee
with the various couplings given by
\vspace*{1ex}
\be
  \begin{array}{|l|c|c|c|}
    \hline & & & \\[-4.5mm]
    S_1S_2S_3S_4             & C_{12}^{(4S)} & C_{13}^{(4S)} & C_{14}^{(4S)} \\
    \hline 
    HHHH                     & 1             & 1             & 1             \\
    \chi\chi\chi\chi         & 1             & 1             & 1             \\
    HH\chi\chi               & 1             & 0             & 0             \\
    HH\phi^+\phi^-           & 1             & 0             & 0             \\
    \chi\chi\phi^+\phi^-     & 1             & 0             & 0             \\
    \phi^+\phi^+\phi^-\phi^- & 0             & 1             & 1             \\
    \hline
  \end{array}
  \vspace*{1ex}
\ee  
The local scalar three- and four-point interactions can be obtained by simply
replacing $\tilde{\Sigma}_H(q^2)$ by $-M_H^2$. 

The non-local action $\SNL^t$ in Eq.~(\ref{sm/NL/top}), finally, modifies 
various three- and four-point interactions between fermions and bosons.
We start with the contribution to the interaction between one scalar particle
and two fermions: 
\be
  \begin{picture}(75,45)(0,-2)
    \DashLine(0,0)(40,0){4}
    \ArrowLine(25,-12)(26,-12)
    \Line(15,-12)(25,-12)
    \ArrowLine(40,0)(60,30)
    \ArrowLine(60,-30)(40,0)
    \GCirc(40,0){3}{0}
    \Text(0,8)[lb]{$S$}
    \Text(0,-8)[lt]{$q$}
    \Text(60,35)[cb]{$\bar{f}_2$}
    \Text(62,12)[lb]{$p'$}
    \Text(60,-35)[ct]{$f_1$}
    \Text(62,-12)[lt]{$p$}
  \end{picture} 
  \hspace*{2ex} 
  :\ \frac{i}{v}\,\biggl[ C_{+}\,\omega_{+}\,\tilde{\Sigma}_t(p^2) 
                        + C_{-}\,\omega_{-}\,\tilde{\Sigma}_t({p'}^2)
                  \biggr],
  \vspace*{7ex}
\ee
with $\omega_{\pm}=(1 \pm \gamma_5)/2$. The various couplings are given by
\vspace*{1ex}
\be
  \begin{array}{|l|c|c|}
    \hline & & \\[-4.5mm]
    Sf_1\bar{f}_2   & C_+       & C_-       \\
    \hline & & \\[-4.5mm] 
    H t\bar{t}      & 1         & 1         \\
    \chi t\bar{t}   & -i        & i         \\
    \phi^- t\bar{b} & -\sqrt{2} & 0         \\
    \phi^+ b\bar{t} & 0         & -\sqrt{2} \\
    \hline
  \end{array}
  \vspace*{1ex}
\ee   
The local top-quark Yukawa interactions can be obtained by simply replacing
$\tilde{\Sigma}_t(q^2)$ by $-m_t$. Owing to the path-ordered exponentials
$U_{1,3}$, the above interaction can be extended by attaching an additional
neutral gauge boson $N_1$:
\begin{eqnarray}
\hspace*{2.5ex}
\begin{picture}(75,45)(0,-2)
{
   \DashLine(0,30)(30,0){4}
   \ArrowLine(7,8)(10,5)
   \Line(7,8)(0,15)
   \Photon(0,-30)(30,0){2}{5}
   \ArrowLine(7,-8)(10,-5)
   \Line(7,-8)(0,-15)
   \ArrowLine(30,0)(60,30)
   \ArrowLine(60,-30)(30,0)
   \GCirc(30,0){3}{0}
   \Text(0,35)[cb]{$S$}
   \Text(-5,19)[rb]{$q$}
   \Text(0,-35)[ct]{$N_1,\,\mu_1$} 
   \Text(-5,-19)[rt]{$q_1$}
   \Text(60,35)[cb]{$\bar{f}_2$}
   \Text(62,11)[lb]{$p'$}
   \Text(60,-35)[ct]{$f_1$}
   \Text(62,-11)[lt]{$p$} 
}
\end{picture}
\hspace*{-0.5ex}
  &:& \hspace*{-1.5ex}\frac{i}{v}\,{\cal G}(N_1)\,
     \sum_{n=0}^{\infty} \frac{1}{n!}\,\tilde{\Sigma}_t^{(n)}(0)\,
     \biggl[ C_{+}\,\omega_{+}\,\tilde{\A}_n^{\mu_1}(q\!-\!p',p|q_1) 
           + C_{-}\,\omega_{-}\,\tilde{\A}_n^{\mu_1}(-p',q\!+\!p|q_1)
     \biggr] 
     \nonumber \\[3mm]
  &\to& \hspace*{-1.5ex}\frac{i}{v}\,{\cal G}(N_1)\,\Biggl[ 
       C_{+}\,\omega_{+}\,(p+p'-q)^{\mu_1}\,
       \frac{\tilde{\Sigma}_t([q-p']^2)-\tilde{\Sigma}_t(p^2)}
            {(q-p')^2-p^2}
       \nonumber \\[3mm]
  & &  \hspace*{-1.5ex}\hphantom{\frac{i}{v}\,{\cal G}Aa}
       +\,C_{-}\,\omega_{-}\,(p+p'+q)^{\mu_1}\,
       \frac{\tilde{\Sigma}_t(p'^{\,2})-\tilde{\Sigma}_t([p+q]^2)}
            {p'^{\,2}-(p+q)^2}      \Biggr].
\end{eqnarray}
The couplings $C_{\pm}$ are the same as without the neutral gauge boson and
the generalized gauge coupling ${\cal G}(N_1)$ is defined as  
\be
  {\cal G}(N_j) = \{-Q_t\,e; -Q_t\,e\,s_w/c_w; g_s\T^{a_j}\}
  \qquad \mbox{for} \qquad
  N_j=\{\gamma;Z;g^{a_j}\},
\ee
with $Q_t=2/3$ denoting the charge of the top-quark in units of $e$. 
The part of $\SNL^t$ that originates from the non-zero vacuum expectation 
value of the Higgs, $\SNL^{t,\,v}$, involves fermions and gauge bosons only
[see Eq.~(\ref{sm/NL/top/nohiggs})]. The corresponding Feynman rules resemble 
the ones derived in Sect.~\ref{sec:top}. For the three- and four-point 
interactions we find
\begin{eqnarray}
\hspace*{2.5ex}
  \begin{picture}(75,45)(0,-2)
    \Photon(0,0)(40,0){2}{5}
    \ArrowLine(25,-12)(26,-12)
    \Line(15,-12)(25,-12)
    \ArrowLine(40,0)(60,30)
    \ArrowLine(60,-30)(40,0)
    \GCirc(40,0){3}{0}
    \Text(0,8)[lb]{$N_1,\,\mu_1$}
    \Text(0,-8)[lt]{$q_1$}
    \Text(60,35)[cb]{$\bar{t}$}
    \Text(62,12)[lb]{$p'$}
    \Text(60,-35)[ct]{$t$}
    \Text(62,-12)[lt]{$p$}
\end{picture}
\hspace*{-0.5ex}
  &:& i\,{\cal G}(N_1)\,
   \sum_{n=0}^{\infty} \frac{1}{n!}\,\tilde{\Sigma}_t^{(n)}(0)\,
   \tilde{\A}_n^{\mu_1}(-p',p|q_1)
     \nonumber \\[3mm]
  &\to& i\,{\cal G}(N_1)\,(p+p')^{\mu_1}\, 
        \frac{\tilde{\Sigma}_t(p'^{\,2})-\tilde{\Sigma}_t(p^2)}{p'^{\,2}-p^2}, 
\end{eqnarray}
\begin{eqnarray}
\hspace*{2.5ex}
\begin{picture}(75,45)(0,-2)
{
   \Photon(0,30)(30,0){2}{5}
   \ArrowLine(7,8)(10,5)
   \Line(7,8)(0,15)
   \Photon(0,-30)(30,0){2}{5}
   \ArrowLine(7,-8)(10,-5)
   \Line(7,-8)(0,-15)
   \ArrowLine(30,0)(60,30)
   \ArrowLine(60,-30)(30,0)
   \GCirc(30,0){3}{0}
   \Text(0,35)[cb]{$N_1,\,\mu_1$}
   \Text(-5,19)[rb]{$q_1$}
   \Text(0,-35)[ct]{$N_2,\,\mu_2$} 
   \Text(-5,-19)[rt]{$q_2$}
   \Text(60,35)[cb]{$\bar{t}$}
   \Text(62,11)[lb]{$p'$}
   \Text(60,-35)[ct]{$t$}
   \Text(62,-11)[lt]{$p$} 
}
\end{picture}
\hspace*{-0.5ex}
  &:& i\,\sum_{n=0}^{\infty} \frac{1}{n!}\,\tilde{\Sigma}_t^{(n)}(0)\,
      \biggl[ {\cal G}(N_1)\,{\cal G}(N_2)\,
              \tilde{\A}_n^{\mu_1\mu_2}(-p',p|q_1,q_2) 
            + (1 \leftrightarrow 2)
      \biggr]
      \nonumber \\[4mm]
  &\to& i\,\frac{{\cal G}(N_1)\,{\cal G}(N_2)}{p'^{\,2}-p^2}\,\Biggl[
        \frac{(p'+p+q_2)^{\mu_1}}{p'^{\,2}-(p+q_2)^2}\,
        \frac{(2p+q_2)^{\mu_2}}{p^2-(p+q_2)^2}\,
        \biggl\{ \Bigl[ p^2-(p+q_2)^2 \Bigr]\,\tilde{\Sigma}_t(p'^{\,2})
        \nonumber \\[4mm]
  & &\hphantom{i\,\frac{{\cal G}(N_1)\,{\cal G}(N_2)}{p'^{\,2}-p^2}\,}
        +\,\Bigl[ p'^{\,2}-p^2 \Bigr]\,\tilde{\Sigma}_t([p+q_2]^2)
        + \Bigl[ (p+q_2)^2-p'^{\,2} \Bigr]\,\tilde{\Sigma}_t(p^2)
        \biggr\}
        \nonumber \\[4mm]
  & &\hphantom{i\,\frac{{\cal G}(N_1)\,{\cal G}(N_2)}{p'^{\,2}-p^2}\,}
        +\,g^{\mu_1\mu_2}\,\Bigl\{ \tilde{\Sigma}_t(p'^{\,2})
                                 - \tilde{\Sigma}_t(p^2) \Bigr\} \Biggr]
        \nonumber \\[4mm]
  & & +\ (N_1,\mu_1,q_1) \leftrightarrow (N_2,\mu_2,q_2).
\end{eqnarray}


\end{document}